\documentclass[twocolumn]{aastex63}
\usepackage{amsmath}
\usepackage{graphicx}
\usepackage{booktabs}

\defcitealias{McGreer18}{M18}
\defcitealias{Kim19}{K19}
\defcitealias{Yang16}{Y16}
\defcitealias{Giallongo19}{G19}

\shorttitle{Quasar Luminosity Function at $z\sim5$}
\shortauthors{Kim et al.}

\begin{document}

\title{The Infrared Medium-deep Survey. VIII. Quasar Luminosity Function at $z\sim5$}
\correspondingauthor{Myungshin Im}
\email{yongjungkim@pku.edu.cn, mim@astro.snu.ac.kr}

\author[0000-0003-1647-3286]{Yongjung Kim}
\altaffiliation{KIAA Fellow}
\affiliation{Kavli Institute for Astronomy and Astrophysics, Peking University, Beijing 100871, P. R. China}
\affiliation{Center for the Exploration of the Origin of the Universe (CEOU), Building 45, Seoul National University, 1 Gwanak-ro, Gwanak-gu, Seoul 08826, Republic of Korea}

\author[0000-0002-8537-6714]{Myungshin Im}
\affiliation{Center for the Exploration of the Origin of the Universe (CEOU), Building 45, Seoul National University, 1 Gwanak-ro, Gwanak-gu, Seoul 08826, Republic of Korea}
\affiliation{Astronomy Program, FPRD, Department of Physics \& Astronomy, Seoul National University, 1 Gwanak-ro, Gwanak-gu, Seoul 08826, Republic of Korea}

\author[0000-0003-4847-7492]{Yiseul Jeon}
\affiliation{FEROKA Inc., 401, Open Innovation Building, Seoul Biohub, 117-3 Hoegi-ro, Dondaemun-gu, Seoul 02455, Republic of Korea}

\author[0000-0002-3560-0781]{Minjin Kim}
\affiliation{Department of Astronomy and Atmospheric Sciences, College of Natural Sciences, Kyungpook National University, Daegu 41566, Republic of Korea}

\author[0000-0002-2548-238X]{Soojong Pak}
\affiliation{Center for the Exploration of the Origin of the Universe (CEOU), Building 45, Seoul National University, 1 Gwanak-ro, Gwanak-gu, Seoul 08826, Republic of Korea}
\affiliation{School of Space Research, Kyung Hee University, 1732 Deogyeong-daero, Giheung-gu, Yongin-si, Gyeonggi-do 17104, Republic of Korea}

\author{Minhee Hyun}
\affiliation{Center for the Exploration of the Origin of the Universe (CEOU), Building 45, Seoul National University, 1 Gwanak-ro, Gwanak-gu, Seoul 08826, Republic of Korea}
\affiliation{Astronomy Program, FPRD, Department of Physics \& Astronomy, Seoul National University, 1 Gwanak-ro, Gwanak-gu, Seoul 08826, Republic of Korea}

\author[0000-0002-0992-5742]{Yoon Chan Taak}
\affiliation{Center for the Exploration of the Origin of the Universe (CEOU), Building 45, Seoul National University, 1 Gwanak-ro, Gwanak-gu, Seoul 08826, Republic of Korea}
\affiliation{Astronomy Program, FPRD, Department of Physics \& Astronomy, Seoul National University, 1 Gwanak-ro, Gwanak-gu, Seoul 08826, Republic of Korea}

\author[0000-0002-2188-4832]{Suhyun Shin}
\affiliation{Center for the Exploration of the Origin of the Universe (CEOU), Building 45, Seoul National University, 1 Gwanak-ro, Gwanak-gu, Seoul 08826, Republic of Korea}
\affiliation{Astronomy Program, FPRD, Department of Physics \& Astronomy, Seoul National University, 1 Gwanak-ro, Gwanak-gu, Seoul 08826, Republic of Korea}

\author{Gu Lim}
\affiliation{Center for the Exploration of the Origin of the Universe (CEOU), Building 45, Seoul National University, 1 Gwanak-ro, Gwanak-gu, Seoul 08826, Republic of Korea}
\affiliation{Astronomy Program, FPRD, Department of Physics \& Astronomy, Seoul National University, 1 Gwanak-ro, Gwanak-gu, Seoul 08826, Republic of Korea}

\author{Gregory S. H. Paek}
\affiliation{Center for the Exploration of the Origin of the Universe (CEOU), Building 45, Seoul National University, 1 Gwanak-ro, Gwanak-gu, Seoul 08826, Republic of Korea}
\affiliation{Astronomy Program, FPRD, Department of Physics \& Astronomy, Seoul National University, 1 Gwanak-ro, Gwanak-gu, Seoul 08826, Republic of Korea}

\author{Insu Paek}
\affiliation{Center for the Exploration of the Origin of the Universe (CEOU), Building 45, Seoul National University, 1 Gwanak-ro, Gwanak-gu, Seoul 08826, Republic of Korea}
\affiliation{Astronomy Program, FPRD, Department of Physics \& Astronomy, Seoul National University, 1 Gwanak-ro, Gwanak-gu, Seoul 08826, Republic of Korea}

\author[0000-0003-4176-6486]{Linhua Jiang}
\affiliation{Kavli Institute for Astronomy and Astrophysics, Peking University, Beijing 100871, P. R. China}
\affiliation{Department of Astronomy, School of Physics, Peking University, Beijing 100871, P. R. China}

\author{Changsu Choi}
\affiliation{Center for the Exploration of the Origin of the Universe (CEOU), Building 45, Seoul National University, 1 Gwanak-ro, Gwanak-gu, Seoul 08826, Republic of Korea}
\affiliation{Astronomy Program, FPRD, Department of Physics \& Astronomy, Seoul National University, 1 Gwanak-ro, Gwanak-gu, Seoul 08826, Republic of Korea}

\author{Jueun Hong}
\affiliation{Center for the Exploration of the Origin of the Universe (CEOU), Building 45, Seoul National University, 1 Gwanak-ro, Gwanak-gu, Seoul 08826, Republic of Korea}
\affiliation{Astronomy Program, FPRD, Department of Physics \& Astronomy, Seoul National University, 1 Gwanak-ro, Gwanak-gu, Seoul 08826, Republic of Korea}

\author{Tae-Geun Ji}
\affiliation{Center for the Exploration of the Origin of the Universe (CEOU), Building 45, Seoul National University, 1 Gwanak-ro, Gwanak-gu, Seoul 08826, Republic of Korea}
\affiliation{School of Space Research, Kyung Hee University, 1732 Deogyeong-daero, Giheung-gu, Yongin-si, Gyeonggi-do 17104, Republic of Korea}

\author[0000-0003-1470-5901]{Hyunsung D. Jun}
\affiliation{Korea Institute for Advanced Study, 85 Hoegi-ro, Dongdaemun-gu, Seoul 02455, Republic of Korea}

\author[0000-0002-8858-3188]{Marios Karouzos}
\affiliation{Nature Astronomy, Springer Nature, 4 Crinan Street, London N1 9XW, UK}

\author[0000-0002-6925-4821]{Dohyeong Kim}
\affiliation{Department of Earth Sciences, Pusan National University, Busan 46241, Republic of Korea}
\affiliation{Kavli Institute for Astronomy and Astrophysics, Peking University, Beijing 100871, P. R. China}
\affiliation{Center for the Exploration of the Origin of the Universe (CEOU), Building 45, Seoul National University, 1 Gwanak-ro, Gwanak-gu, Seoul 08826, Republic of Korea}

\author[0000-0001-5120-0158]{Duho Kim}
\affiliation{Korea Astronomy and Space Science Institute, Daejeon 34055, Republic of Korea}

\author[0000-0002-1710-4442]{Jae-Woo Kim}
\affiliation{Korea Astronomy and Space Science Institute, Daejeon 34055, Republic of Korea}

\author[0000-0002-1418-3309]{Ji Hoon Kim}
\affiliation{Metaspace Inc., 36 Nonhyeon-ro, Gangnam-gu, Seoul 06312, Republic of Korea}

\author{Hye-In Lee}
\affiliation{Center for the Exploration of the Origin of the Universe (CEOU), Building 45, Seoul National University, 1 Gwanak-ro, Gwanak-gu, Seoul 08826, Republic of Korea}
\affiliation{School of Space Research, Kyung Hee University, 1732 Deogyeong-daero, Giheung-gu, Yongin-si, Gyeonggi-do 17104, Republic of Korea}

\author[0000-0001-5342-8906]{Seong-Kook Lee}
\affiliation{Center for the Exploration of the Origin of the Universe (CEOU), Building 45, Seoul National University, 1 Gwanak-ro, Gwanak-gu, Seoul 08826, Republic of Korea}
\affiliation{Astronomy Program, FPRD, Department of Physics \& Astronomy, Seoul National University, 1 Gwanak-ro, Gwanak-gu, Seoul 08826, Republic of Korea}

\author[0000-0002-8292-2556]{Won-Kee Park}
\affiliation{Korea Astronomy and Space Science Institute, Daejeon 34055, Republic of Korea}

\author{Yongmin Yoon}
\affiliation{Korea Institute for Advanced Study, 85 Hoegi-ro, Dongdaemun-gu, Seoul 02455, Republic of Korea}

\author{Seoyeon Byeon}
\affiliation{Center for the Exploration of the Origin of the Universe (CEOU), Building 45, Seoul National University, 1 Gwanak-ro, Gwanak-gu, Seoul 08826, Republic of Korea}
\affiliation{School of Space Research, Kyung Hee University, 1732 Deogyeong-daero, Giheung-gu, Yongin-si, Gyeonggi-do 17104, Republic of Korea}

\author{Sungyong Hwang}
\affiliation{Center for the Exploration of the Origin of the Universe (CEOU), Building 45, Seoul National University, 1 Gwanak-ro, Gwanak-gu, Seoul 08826, Republic of Korea}
\affiliation{Astronomy Program, FPRD, Department of Physics \& Astronomy, Seoul National University, 1 Gwanak-ro, Gwanak-gu, Seoul 08826, Republic of Korea}

\author{Joonho Kim}
\affiliation{Center for the Exploration of the Origin of the Universe (CEOU), Building 45, Seoul National University, 1 Gwanak-ro, Gwanak-gu, Seoul 08826, Republic of Korea}
\affiliation{Astronomy Program, FPRD, Department of Physics \& Astronomy, Seoul National University, 1 Gwanak-ro, Gwanak-gu, Seoul 08826, Republic of Korea}

\author{Sophia Kim}
\affiliation{Center for the Exploration of the Origin of the Universe (CEOU), Building 45, Seoul National University, 1 Gwanak-ro, Gwanak-gu, Seoul 08826, Republic of Korea}
\affiliation{Astronomy Program, FPRD, Department of Physics \& Astronomy, Seoul National University, 1 Gwanak-ro, Gwanak-gu, Seoul 08826, Republic of Korea}

\author{Woojin Park}
\affiliation{Center for the Exploration of the Origin of the Universe (CEOU), Building 45, Seoul National University, 1 Gwanak-ro, Gwanak-gu, Seoul 08826, Republic of Korea}
\affiliation{School of Space Research, Kyung Hee University, 1732 Deogyeong-daero, Giheung-gu, Yongin-si, Gyeonggi-do 17104, Republic of Korea}




\begin{abstract}
Faint $z\sim5$ quasars with $M_{1450}\sim-23$ mag are known to be the potentially important contributors to the ultraviolet ionizing background in the post-reionization era. However, their number density has not been well determined, making it difficult to assess their role in the early ionization of the intergalactic medium (IGM). In this work, we present the updated results of our $z\sim5$ quasar survey using the Infrared Medium-deep Survey (IMS), a near-infrared imaging survey covering an area of 85 deg$^{2}$.
From our spectroscopic observations with the Gemini Multi-Object Spectrograph (GMOS) on the Gemini-South 8 m Telescope, we discovered eight new quasars at $z\sim5$ with $-26.1\leq M_{1450} \leq -23.3$. 
Combining our IMS faint quasars ($M_{1450}>-27$ mag) with the brighter Sloan Digital Sky Survey (SDSS) quasars ($M_{1450}<-27$ mag), we derive the $z\sim5$ quasar luminosity function (QLF) without any fixed parameters down to the magnitude limit of $M_{1450}=-23$ mag.
We find that the faint-end slope of the QLF is very flat ($\alpha=-1.2^{+1.4}_{-0.6}$), with a characteristic luminosity of $M^{*}_{1450}=-25.8^{+1.4}_{-1.1}$ mag. The number density of $z\sim5$ quasars from the QLF gives an ionizing emissivity at $912~\rm\AA$ of $\epsilon_{912}=(3.7$--$7.1)\times10^{23}$ erg s$^{-1}$ Hz$^{-1}$ Mpc$^{-3}$ and an ionizing photon density of $\dot{n}_{\rm ion}=(3.0$--$5.7)\times10^{49}$ Mpc$^{-3}$ s$^{-1}$. These results imply that quasars are responsible for only 10-20\% (up to 50\% even in the extreme case) of the photons required to completely ionize the IGM at $z\sim5$, disfavoring the idea that quasars alone could have ionized the IGM at $z\sim5$.
\end{abstract}



\section{INTRODUCTION} \label{sec:introduction}

Quasars are known as key objects for understanding the universe along cosmic time, especially at high redshifts ($z\gtrsim5$) where galaxies are hard to detect.
The Sloan Digital Sky Survey (SDSS) pioneered the identification of quasars at $z\sim6$, providing the first glimpse of the reionization process in the early universe (e.g., \citealt{Fan01,Fan06,Jiang16}).
As new surveys have explored wider areas and deeper limits, the number of known quasars at high redshifts has steadily increased in the last few decades.
Along with the record holder ULAS J1345$+$0928 at an extreme redshift of $z=7.54$ \citep{Banados18}, hundreds of high-redshift quasars are being discovered by various surveys \citep{Willott10b,Mortlock11,McGreer13,McGreer18,Venemans13,Venemans15a,Venemans15b,
Banados14,Banados16,Kashikawa15,Kim15,Kim19,
Wu15,Jiang16,Matsuoka16,Matsuoka18a,Matsuoka19,Wang16,Wang17,Wang18,Wang19,
Yang16,Yang17,Yang19a,Yang19b,Yang20,Jeon17,Reed17,Shin20}.
This large sample of distant quasars enables us to construct quasar luminosity functions (QLFs) at high redshifts, which are important for investigating the contribution of high-redshift quasars to the cosmic ultraviolet (UV) ionizing background and the evolution of quasar populations along the redshift.

\begin{figure}[t]
\centering
\epsscale{1.2}
\plotone{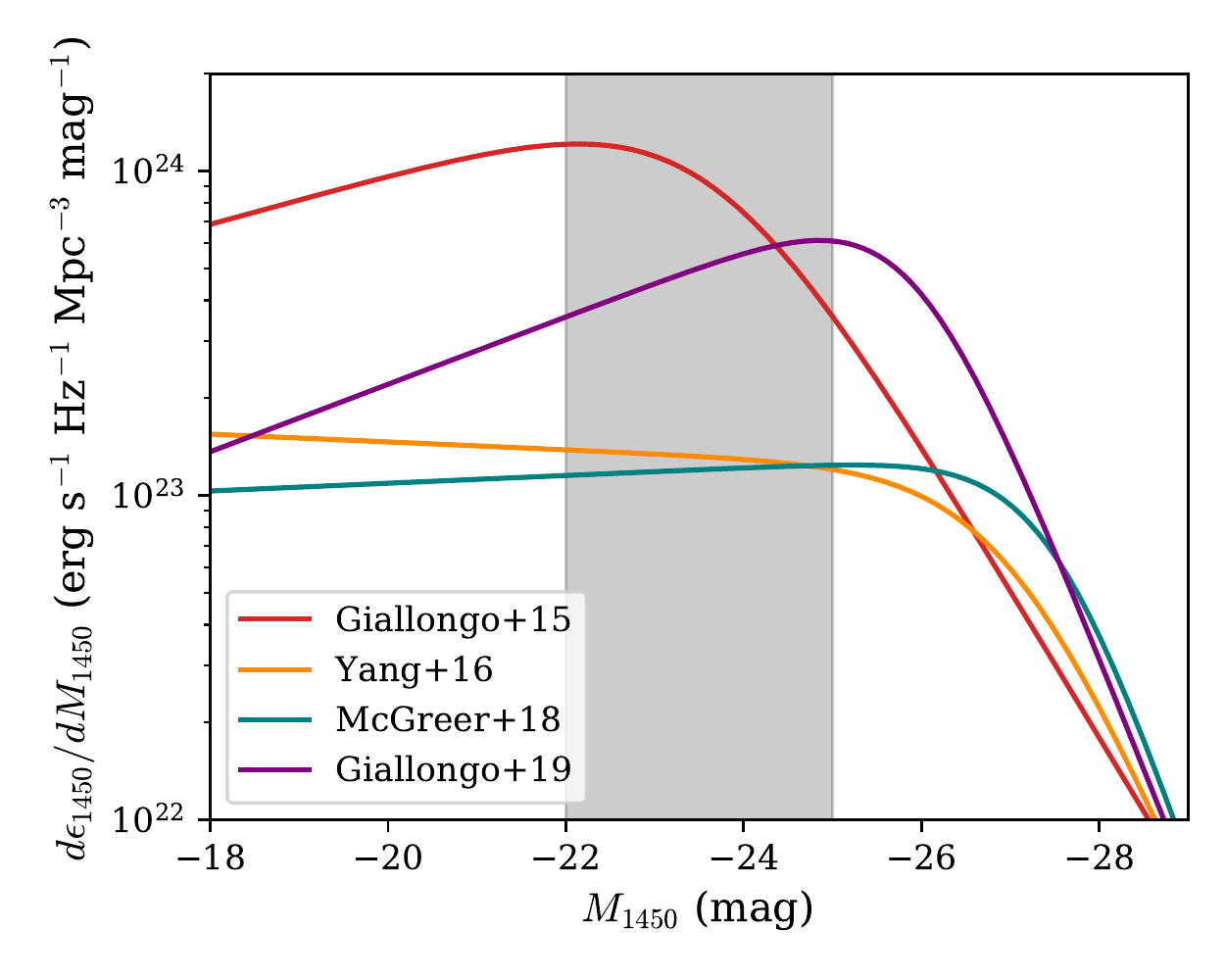}
\caption{
Differential contributions to $z\sim5$ quasar emissivity, calculated from the QLFs in the literature: \citeauthor{Giallongo15} (\citeyear{Giallongo15}; red), \citeauthor{Yang16} (\citeyear{Yang16}; orange), \citeauthor{McGreer18} (\citeyear{McGreer18}; teal), and \citeauthor{Giallongo19} (\citeyear{Giallongo19}; purple).
The gray shaded region shows the magnitude range in which quasars make a considerable contribution to total quasar emissivity.
\label{fig:de}}
\end{figure}

Since the cosmic reionization at $6<z<8.8$ (e.g., \citealt{McGreer15,Planck16,Madau17,Davies18,Mason18,Mason19}), most of the hydrogen atoms in the intergalactic medium (IGM) have remained ionized until $z=0$.
It has been debated whether high-redshift quasars that are bright in UV wavelengths can provide the enormous quantity of ionizing photons required to keep hydrogen atoms ionized.
The number of ionizing photons from quasars can be calculated from their UV emissivity, proportional to both luminosity and quasar number density.
According to recently derived QLFs, the emissivity critically depends on the faint quasars at $M_{1450}\sim -23.5$ mag (absolute magnitude at $1450~\rm\AA$ in the rest frame).
Figure \ref{fig:de} shows the UV emissivity from the quasars for different magnitude bins.
There are differences between QLFs, but $M_{1450}\sim-23.5$ mag quasars make a considerable contribution to the total quasar emissivity.
However, it has been difficult to locate them in the past owing to observational limitations, such as the shallow imaging depths of large-area surveys or the small survey areas of deep surveys.
As a result, past determination of the QLF at $M_{1450}\sim-23.5$ mag often relied on uncertain extrapolations of the QLF \citep{Giallongo15,Yang16}.

Owing to new deep and wide-area optical/near-infrared (NIR) surveys, tens of such faint quasars are now being uncovered (e.g., \citealt{Kashikawa15,Kim15,Kim19,Matsuoka16,Matsuoka18a,Matsuoka19,Akiyama18,McGreer18,Shin20}).
However, the contribution of quasars to the IGM-ionizing UV photons remains controversial.
On one hand, the results of optical/NIR wide-area surveys suggest that high-redshift quasars are not the main provider of UV photons, contributing $<50$\% of the required photons at $z\sim5$ \citep{Yang16,McGreer18,Shin20} and $<10$\% at $z\sim6$  \citep{Willott10b,Kashikawa15,Kim15,Matsuoka18b}.
On the other hand, from very deep surveys covering small areas ($\lesssim 4$ deg$^{2}$), other studies have revealed a large number of faint quasars at high redshifts, supporting a significant contribution of quasars to the IGM ionizing background \citep{Glikman11,Boutsia18,Grazian20}.
In particular, for X-ray-selected quasars with $M_{1450} \geq -22$ mag, \citeauthor{Giallongo19} (\citeyear{Giallongo19}; hereafter \citetalias{Giallongo19}) have suggested a substantial contribution of quasars ($\gtrsim 50$\%, depending on $z$) to the IGM ionizing background (see also \citealt{Giallongo15}).

At $z\sim5$, \citeauthor{McGreer18} (\citeyear{McGreer18}, hereafter \citetalias{McGreer18}), performed a quasar search using the data from the Canada--France--Hawaii Telescope Legacy Survey (CFHTLS; \citealt{Hudelot12}).
They derived the QLF at $4.7<z<5.4$, combining the bright SDSS quasar sample of \cite{McGreer13} and 25 newly discovered CFHTLS quasars with $M_{1450}\lesssim-23$ mag.
Their QLF, in the form of a double power-law function, has a faint-end slope of $-1.97$ and a low number density, implying that quasars are unlikely to be the main UV ionizing source at $z\sim5$.
Due to the lack of faint quasars identified with spectroscopy, however, they fixed the bright-end slope of the QLF to $-4.0$, which may have given biased interpretations of the quasar population (and corresponding emissivity) with underestimated uncertainties \citep{Kulkarni19}.
To date, only \citeauthor{Yang16} (\citeyear{Yang16}, hereafter \citetalias{Yang16}), have derived the $z\sim5$ QLF without fixed parameters, including the quasar sample of \cite{McGreer13}.
However, most of the quasars they used are $M_{1450}<-24$ mag, which might not be sufficient to derive the QLF precisely.
In fact, their best-fit parameters show discontinuity along the redshift, compared to the recently reported QLFs at $z\sim4$ \citep{Akiyama18} and 6 \citep{Matsuoka18b}, which are from a wide range of luminosities $(-30 \lesssim M_{1450} \lesssim -22)$ over a large survey area without any fixed parameters.

Recently, our group has been performing a $z\sim5$ quasar survey (\citealt{Kim19}, hereafter \citetalias{Kim19}), with the Infrared Medium-deep Survey (IMS; M. Im et al. 2020, in preparation), for which NIR imaging data were obtained with the Wide Field Camera (WFCam; \citealt{Casali07}) on the United Kingdom Infrared Telescope (UKIRT).
The data cover an area of $\sim100$ deg$^{2}$ over several extragalactic fields and reach $5\sigma$ depths of $J\sim23$ mag.
We combined the IMS data with the optical data from the CFHTLS \citep{Hudelot12} and additional NIR data from the Deep eXtragalactic Survey (DXS) of the UKIRT Infrared Deep Sky Survey (UKIDSS; \citealt{Lawrence07}).
Adopting the medium-band color selections that are known to improve the high-redshift quasar selection efficiency \citep{Jeon16}, we spectroscopically identified 13 faint quasars at $z\sim5$ in the past, of which 10 were newly discovered.
Increasing the number of spectroscopically identified quasars enabled us to refine and constrain the faint-end slope of the $z\sim5$ QLF.
In this study, we have added more samples and derived the QLF at $z\sim5$.

The remainder of this paper is organized as follows.
We present the newly discovered $z\sim5$ quasars from our survey in Section \ref{sec:newquasars}, including a summary of our survey progress.
In Section \ref{sec:qlf}, we describe the derivation of the QLF using the IMS $z\sim5$ quasars.
We discuss the implications of the $z\sim5$ QLF regarding the cosmic ionizing background in Section \ref{sec:discussion}.
We adopt the cosmological parameters of $\Omega_{m}=0.3$, $\Omega_{\Lambda}=0.7$, and $H_{0}=70$ km s$^{-1}$ Mpc$^{-1}$, which are supported by observations in recent decades (e.g., \citealt{Im97,Planck18}).
All magnitudes given in this paper are in the AB system.

\begin{figure}[t]
\centering
\epsscale{1.1}
\plotone{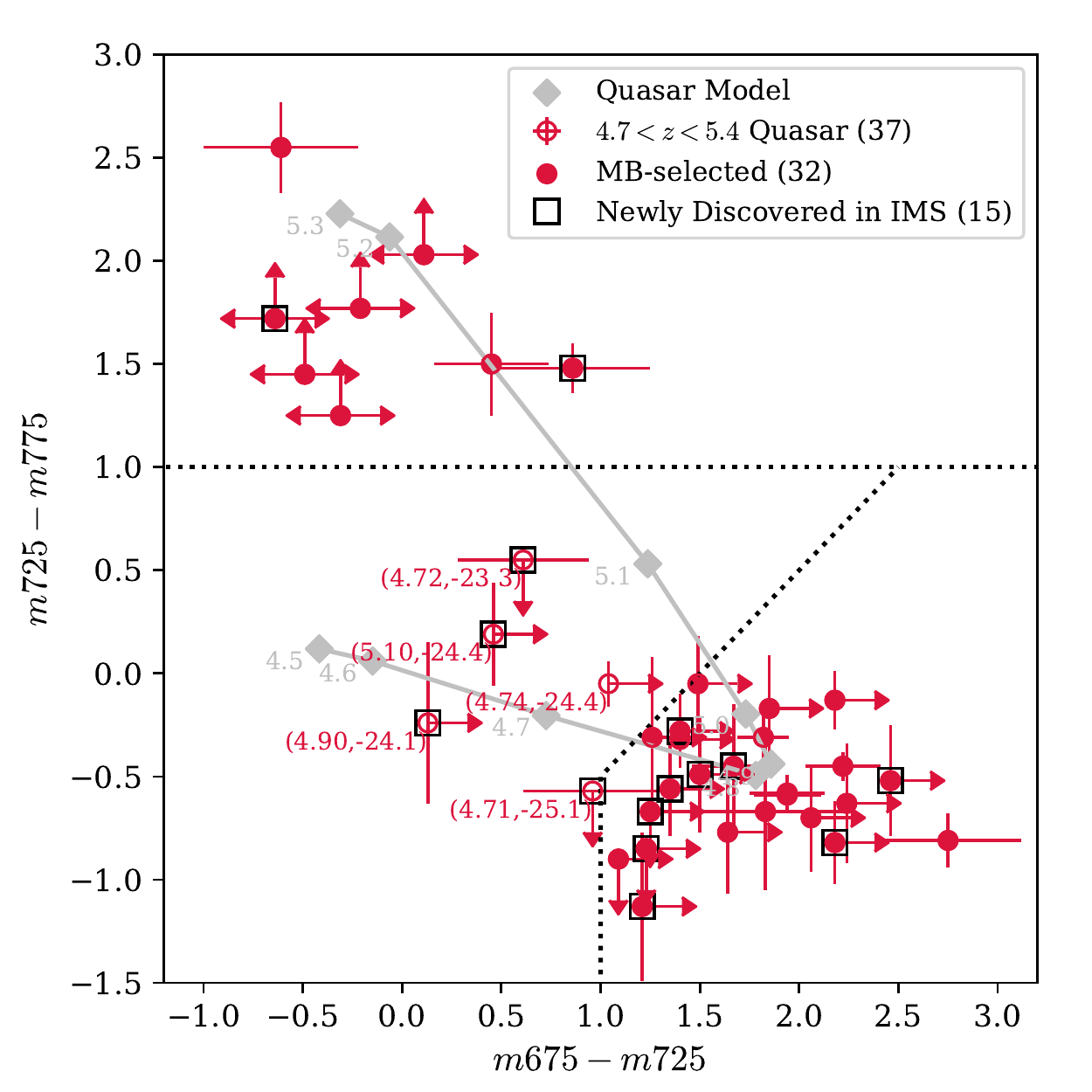}
\caption{
Medium-band color--color diagram of the quasars observed in all three medium bands ($m675$/$m725$/$m775$).
The circles denote 37 quasars that are spectroscopically identified at $4.7<z<5.4$, while the 32 quasars satisfying our color selection criteria (dotted lines) are highlighted as filled circles.
The arrows indicate the upper limits on the magnitude according to the detection limits of their medium-band images.
For the five quasars outside the selection boxes, their ($z$, $M_{1450}$) values are denoted at the lower left of each symbol.
The newly spectroscopically confirmed quasars by us (\citetalias{Kim19} and this work) are highlighted with squares.
The gray diamonds with lines denote the quasar color tracks from $z=4.5$ to 5.3.
\label{fig:ccd}}
\end{figure}

\section{NEW IMS QUASARS AT $\lowercase{z}\sim5$}\label{sec:newquasars}

Here we summarize the IMS $z\sim5$ quasar survey, described in detail in \citetalias{Kim19},
with the newly discovered quasars at $z\sim5$.

\subsection{Initial Broadband Selection\label{sec:bbsel}}

The initial photometric dataset is a combination of the optical imaging data from the CFHTLS Wide Survey and the NIR imaging data from IMS and DXS, covering four extragalactic fields: XMM-Large Scale Structure survey region (XMM-LSS), CFHTLS Wide survey second region (CFHTLS-W2), Extended Groth Strip (EGS), and Small Selected Area 22h (SA22).
The average 5$\sigma$ limiting magnitudes for point sources are $u'=26.1$, $g'=26.4$, $r'=25.9$, $i'=25.6$, $z'=24.6$, and $J=22.9$ mag.
Note that the optical photometric system here is based on the SDSS filter system transformed from those of the CFHTLS\footnote{\url{http://www.cadc-ccda.hia-iha.nrc-cnrc.gc.ca/en/megapipe/docs/filtold.html}}.
The total survey area used for the $z\sim5$ quasar search covers $\sim85$ deg$^{2}$ of the sky, calculated from the full overlapping region of the surveys. 
For the $i'$-band detected sources, we selected $z\sim5$ quasar candidates using the broadband color selection criteria: (1) $i' < 23$, (2) S/N ($u') < 2.5$, (3) $g'-r'>1.8$ or S/N ($g')<3.0$, (4) $r'-i'>1.2$, (5) $i'-z'<0.625\times((r'-i')-1.0)$, (6) $i'-z'<0.55$, (7) $i'-J<( (r'-i') -1.0) + 0.56$.
In the visual inspection stage, seven sources that are unusually elongated or extended were rejected, giving the final 69 $z\sim5$ quasar candidates\footnote{The total number of broadband-selected candidates is decreased from 70 in \citetalias{Kim19} because of the miscoded procedure. Four candidates, including two spectroscopically identified non-quasars,  are rejected, and three candidates without medium-band observations are included instead.}.

\subsection{SQUEAN Follow-up Imaging in Medium Bands}

To narrow down the number of plausible $z\sim5$ quasar candidates, follow-up observations of these broadband-selected candidates were carried out in medium bands with the SED Camera for Quasars in Early Universe (SQUEAN; \citealt{Choi15,Kim16}) on the Otto Struve 2.1 m Telescope at McDonald Observatory.
At $z\sim5$, the Ly$\alpha~\lambda1216$ line is located at $\sim7300~\rm\AA$, so we have obtained $m675$, $m725$, and $m775$ images, where the filter names are chosen by combining ``$m$'' for medium-band and the number of the central wavelength of the filter in nanometers.
These 50 nm width filters finely sample the redshifted Ly$\alpha$ line to improve quasar identification and determine their redshifts to nearly 1\% accuracy (\citealt{Jeon16}; \citetalias{Kim19}).
The total exposure times in medium bands are proportional to the $i'$-band magnitude of each target, as are the resultant imaging depths.
These inhomogeneous imaging depths between targets are important when we calculate the medium-band selection functions in Section \ref{sec:selfun}.
Most of the broadband-selected candidates were observed in the $m725$ and $m775$ bands (63/69), while 50 were also observed in the $m675$ band.
We introduced the medium-band color selection criteria of \cite{Jeon16}: (1) $m675-m725 > 1.0$ and $m675-m725 > m725-m775+1.5$ ($4.7<z<5.1$), (2) $m725-m775>1$ ($5.1<z<5.5$).
Among the 50 medium-band observed candidates, 33 satisfy the criteria.
In Figure \ref{fig:ccd}, we plot a $m675-m725$ vs $m725-m775$ diagram with the medium-band color criteria shown as dotted lines.
For simplicity, we only present the 37 spectroscopically identified $4.7<z<5.4$ quasars that are also observed in the three medium bands.
There are five quasars excluded by our selection (open circles), possibly due to their $m675$ imaging depths being shallower than expected for a given $i'$-band magnitude as described in Section \ref{sec:selfun}, and/or their marginal redshift of $z\simeq4.7$.

\begin{figure}
\centering
\epsscale{1.2}
\plotone{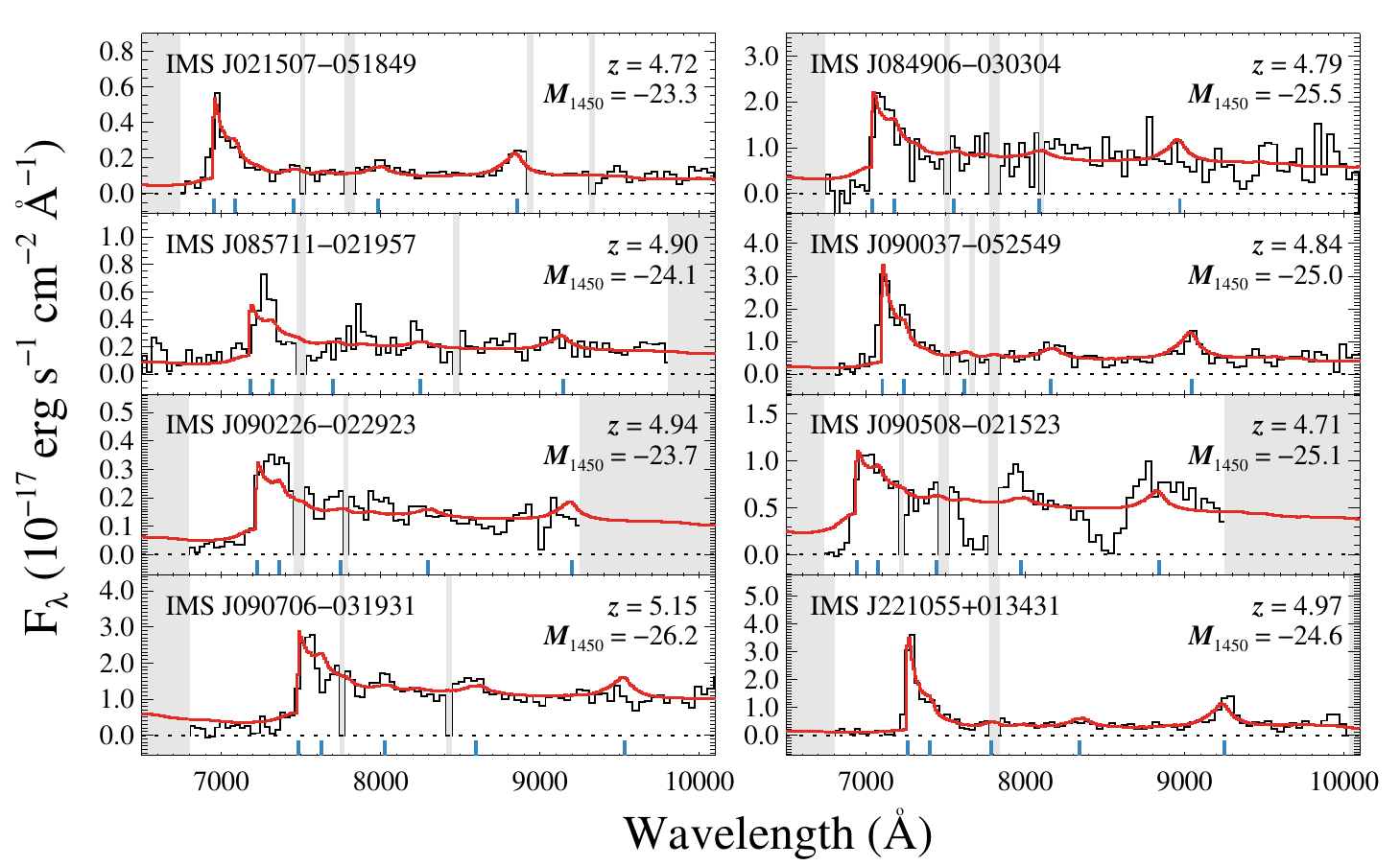}
\caption{
GMOS optical spectra of the newly discovered IMS $z\sim5$ quasars.
The black solid lines represent the binned spectra according to spectral resolution, while the red solid lines are the best-fit models.
The blue marks are the wavelengths of possible quasar emission lines: Ly$\beta$, Ly$\alpha$, \ion{N}{5}, \ion{O}{1}, \ion{Si}{4}, and \ion{C}{4}, from short to long wavelengths.
The shaded regions represent the bad columns, hot pixels, CCD gaps, or wavelength range that are not covered by the observational configuration. 
\label{fig:spec}}
\end{figure}

\begin{deluxetable*}{lccccccccc}
\tabletypesize{\scriptsize}
\tablecaption{Broadband and Medium-band Photometry of New IMS $z\sim5$ Quasars \label{tbl:broad}}
\tablewidth{0pt}
\tablehead{
\colhead{ID} & \colhead{R.A.} & \colhead{Decl.} & \colhead{$r'$} & \colhead{$i'$} & \colhead{$z'$} & \colhead{$J$} & \colhead{$m675$} & \colhead{$m725$} & \colhead{$m775$} \\
\colhead{} & (J2000) & (J2000) & (mag) & (mag) & (mag) & (mag) & (mag) & (mag) & (mag)
}
\startdata
IMS J021507$-$051849 & 02:15:07.31 & $-$05:18:49.2 & $24.16\pm0.08$ & $22.87\pm0.03$ & $22.65\pm0.06$ & $>22.65$ & $23.99\pm0.25$ & $23.38\pm0.21$ & $>22.83$ \\
IMS J084906$-$030304 & 08:49:06.33 & $-$03:03:03.9 & $22.31\pm0.03$ & $20.87\pm0.01$ & $20.67\pm0.02$ & $21.21\pm0.07$ & $>22.65$ & $21.25\pm0.12$ & $21.53\pm0.13$ \\
IMS J085711$-$021957 & 08:57:10.66 & $-$02:19:57.3 & $23.85\pm0.07$ & $22.28\pm0.02$ & $22.40\pm0.08$ & $22.30\pm0.16$ & $>23.01$ & $22.88\pm0.20$ & $23.12\pm0.34$ \\
IMS J090037$-$052549 & 09:00:36.89 & $-$05:25:49.4 & $22.66\pm0.03$ & $21.02\pm0.01$ & $21.15\pm0.02$ & $21.20\pm0.10$ & $>23.15$ & $20.97\pm0.08$ & $21.79\pm0.18$ \\
IMS J090226$-$022923 & 09:02:25.81 & $-$02:29:22.8 & $24.52\pm0.06$ & $22.67\pm0.02$ & $22.26\pm0.05$ & $22.61\pm0.32$ & ... & $22.77\pm0.14$ & $22.71\pm0.15$ \\
IMS J090508$-$021523 & 09:05:08.10 & $-$02:15:23.4 & $22.93\pm0.04$ & $21.44\pm0.01$ & $21.07\pm0.02$ & $20.92\pm0.07$ & $22.79\pm0.30$ & $21.83\pm0.18$ & $>22.40$ \\
IMS J090706$-$031931 & 09:07:06.12 & $-$03:19:30.6 & $22.30\pm0.02$ & $20.62\pm0.01$ & $20.20\pm0.01$ & $20.34\pm0.04$ & $22.61\pm0.37$ & $21.75\pm0.11$ & $20.27\pm0.05$ \\
IMS J221055$+$013430 & 22:10:54.87 & $+$01:34:30.5 & $22.98\pm0.03$ & $21.32\pm0.01$ & $21.27\pm0.03$ & $21.88\pm0.08$ & $>23.65$ & $21.19\pm0.19$ & $21.71\pm0.19$ \\
\enddata
\tablecomments{All magnitudes are given in the AB system. The detection limits ($\sim3\sigma$) are given as the upper limits for objects that were undetected or fainter than the limit.
}
\end{deluxetable*}

\begin{deluxetable*}{lccc}
\tabletypesize{\scriptsize}
\tablecaption{GMOS-S Observations of New IMS $z\sim5$ Quasars \label{tbl:spec}}
\tablewidth{0pt}
\tablehead{
\colhead{ID} & \colhead{Date} & \colhead{Exposure time (s)} & \colhead{Seeing (arcsec)}
}
\startdata
IMS J021507$-$051849 	& 2018 Sep 17	& 4356	& 0.8 \\
IMS J084906$-$030304 	& 2019 Jan 2	& 242	& 0.8 \\
IMS J085711$-$021957 	& 2019 Jan 11	& 605	& 0.8 \\
IMS J090037$-$052549 	& 2019 Jan 2	& 242	& 0.8 \\
IMS J090226$-$022923 	& 2019 Feb 24	& 4356	& 0.8 \\
IMS J090508$-$021523 	& 2019 Feb 14	& 968	& 1.1 \\
IMS J090706$-$031931 	& 2019 Jan 3	& 218	& 1.1 \\
IMS J221055$+$013430 	& 2018 Sep 17	& 484	& 0.5 \\
\enddata
\end{deluxetable*}

\begin{deluxetable*}{lccc}
\tabletypesize{\scriptsize}
\tablecaption{Spectral Fitting Results for New IMS $z\sim5$ Quasars\label{tbl:quantity}}
\tablewidth{0pt}
\tablehead{
\colhead{ID} & \colhead{$z$} & \colhead{$M_{1450}$} & \colhead{$\log$ EW$_{\rm Ly\alpha+NV}$}
}
\startdata
IMS J021507$-$051849 	& $4.716^{+0.003}_{-0.003}$	& $-23.28^{+0.19}_{-0.17}$	& $1.78^{+0.16}_{-0.17}$ \\
IMS J084906$-$030304 	& $4.790^{+0.095}_{-0.225}$	& $-25.50^{+0.75}_{-0.45}$	& $1.50^{+0.49}_{-1.50}$ \\
IMS J085711$-$021957 	& $4.904^{+0.003}_{-0.003}$	& $-24.10^{+0.06}_{-0.06}$	& $1.43^{+0.11}_{-0.12}$ \\
IMS J090037$-$052549 	& $4.840^{+0.004}_{-0.023}$	& $-25.04^{+0.36}_{-0.23}$	& $1.90^{+0.21}_{-0.21}$ \\
IMS J090226$-$022923 	& $4.939^{+0.003}_{-0.020}$	& $-23.72^{+0.14}_{-0.09}$	& $1.38^{+0.17}_{-0.27}$ \\
IMS J090508$-$021523 	& $4.707^{+0.031}_{-0.022}$	& $-25.05^{+0.24}_{-0.22}$	& $1.29^{+0.32}_{-0.59}$ \\
IMS J090706$-$031931 	& $5.154^{+0.017}_{-0.004}$	& $-26.20^{+0.07}_{-0.03}$	& $1.39^{+0.14}_{-0.14}$ \\
IMS J221055$+$013430 	& $4.968^{+0.004}_{-0.015}$	& $-24.60^{+0.57}_{-0.35}$	& $2.12^{+0.27}_{-0.27}$ \\
\enddata
\end{deluxetable*}

\subsection{GMOS Spectroscopy}

We obtained the optical spectra of the plausible candidates, which satisfied either the above broadband or medium-band color selection criteria, with the Gemini Multi-Object Spectrograph (GMOS; \citealt{Hook04}) on the Gemini-South 8 m Telescope on 2018 Sep 17 (PID: GS-2018B-Q-217), 2019 Jan 2--11, and 2019 Feb 14--24 (PID: GS-2019A-Q-218), under the seeing condition of $\lesssim1\farcs0$.
The observing configurations were set to increase the signal-to-noise ratio (S/N) of the spectrum; we used the Nod \& Shuffle mode for sky subtraction, an R150\_G5326 grating with a resolution of $R\sim315$, and spectral/spatial binning of $4\times4$.
In total, we obtained the spectra of ten candidates observed in at least two medium bands.
For spectral data reduction, we followed the procedure in \citetalias{Kim19} which is summarized here briefly.
We used the Gemini IRAF package for the basic reductions: the bias subtraction, flat-fielding, sky-lines subtraction, wavelength calibration with CuAr arc lines, and flux calibration with standard stars.
After the 1D extraction with an aperture with a diameter of 1\farcs0, the overall flux of each spectrum was scaled with the $i'$-band magnitude of each target.
To maximize S/N, we binned the spectra along the wavelength accounting of the instrumental resolution of $\sim300$, using the inverse-variance weighting method (e.g., \citealt{Kim18}).

Through these spectroscopic observations, eight $z\sim5$ quasars were newly discovered, the optical spectra of which are shown in Figure \ref{fig:spec}.
All of them show strong Ly$\alpha$ emission lines with sharp breaks, and some also show other possible emission lines like \ion{C}{4}, while there are broad absorption lines (BAL) in the spectrum of IMS J090508$-$021523.

We list the broadband and medium-band photometry of the new IMS $z\sim5$ quasars in Table \ref{tbl:broad}, and their spectroscopic observations are summarized in Table \ref{tbl:spec}.
The other two candidates identified as non-quasars are listed in Appendix \ref{sec:addspec}.

\subsection{Spectral Fitting\label{sec:specfit}}

We measured the spectral properties of the eight new quasars by fitting their spectra with a refined version of our high-redshift quasar model (\citetalias{Kim19}).
The model is based on the composite quasar spectrum of SDSS quasars \cite{Vanden01}.
Since no IGM correction has been applied in this spectrum, we replaced the spectrum at $\lambda<1216~\rm\AA$ to that of \cite{Lusso15}.
Note that the two spectra were normalized at $1450~\rm\AA$ before merging.
For this composite spectrum, the IGM attenuation by neutral hydrogen was applied in the form of a function of redshift \citep{Madau96}.
The model has four parameters: redshift ($z$), absolute magnitude ($M_{1450}$), continuum slope ($\alpha_{\lambda}$), and equivalent width (EW) of Ly$\alpha~\lambda1216$ and \ion{N}{5} $\lambda1250$ (EW$_{\rm Ly\alpha+NV}$).
The SDSS composite spectrum is first decomposed into the emission line component and the continuum component by fitting a power-law continuum to it.
The slope of the continuum component is allowed to vary to a given $\alpha_{\lambda}$ by multiplying a factor of $(\lambda_{\rm rest}/1000~{\rm\AA})^{\alpha_{\lambda}+1.54}$.
EW$_{\rm Ly\alpha+NV}$, estimated from the composite continuum-subtracted flux at $1160\leq \lambda_{\rm rest}~({\rm\AA})\leq 1290$, is also scaled with $(\lambda_{\rm rest}/1290~{\rm\AA})^{p}$, where $p$ is the appropriate power to adjust the EV$_{\rm Ly\alpha+NV}$ value of the model to an arbitrary one.
In the original \citetalias{Kim19} model, only EW of Ly$\alpha+$\ion{N}{5} was allowed to scale.
On the other hand, in the refined \citetalias{Kim19} model, we also adjusted the other emission lines, such as \ion{C}{4} $\lambda1549$ and \ion{C}{3}] $\lambda1909$, by scaling fluxes over continuum with EW$_{\rm Ly\alpha+NV}$.
This is because EWs of these emission lines are known to scale with EW$_{\rm Ly\alpha+NV}$ \citep{Dietrich02} and have EW distributions with scatters similar to that of Ly$\alpha+$\ion{N}{5} (0.2-0.3 dex; \citealt{Diamond09,Shen11}).

We found the best-fit model for each quasar spectrum by finding the minimum $\chi^{2}$ values.
Considering the narrow wavelength coverage of our spectra, we fixed $\alpha_{\lambda}$ to $-1.54$, same as \citetalias{Kim19}.
In Figure \ref{fig:spec}, the best-fit models for the eight new quasars are shown as the red solid lines, and we listed the best-fit values in Table \ref{tbl:quantity}.
For IMS J090508$-$021523, a BAL quasar, the fluxes at BAL wavelengths are excluded from the fitting.
Note that this model is also used to determine the selection functions in Section \ref{sec:selfun}.

\section{QUASAR LUMINOSITY FUNCTION AT $\lowercase{z}\sim5$}\label{sec:qlf}

\begin{figure*}
\centering
\epsscale{1.0}
\plotone{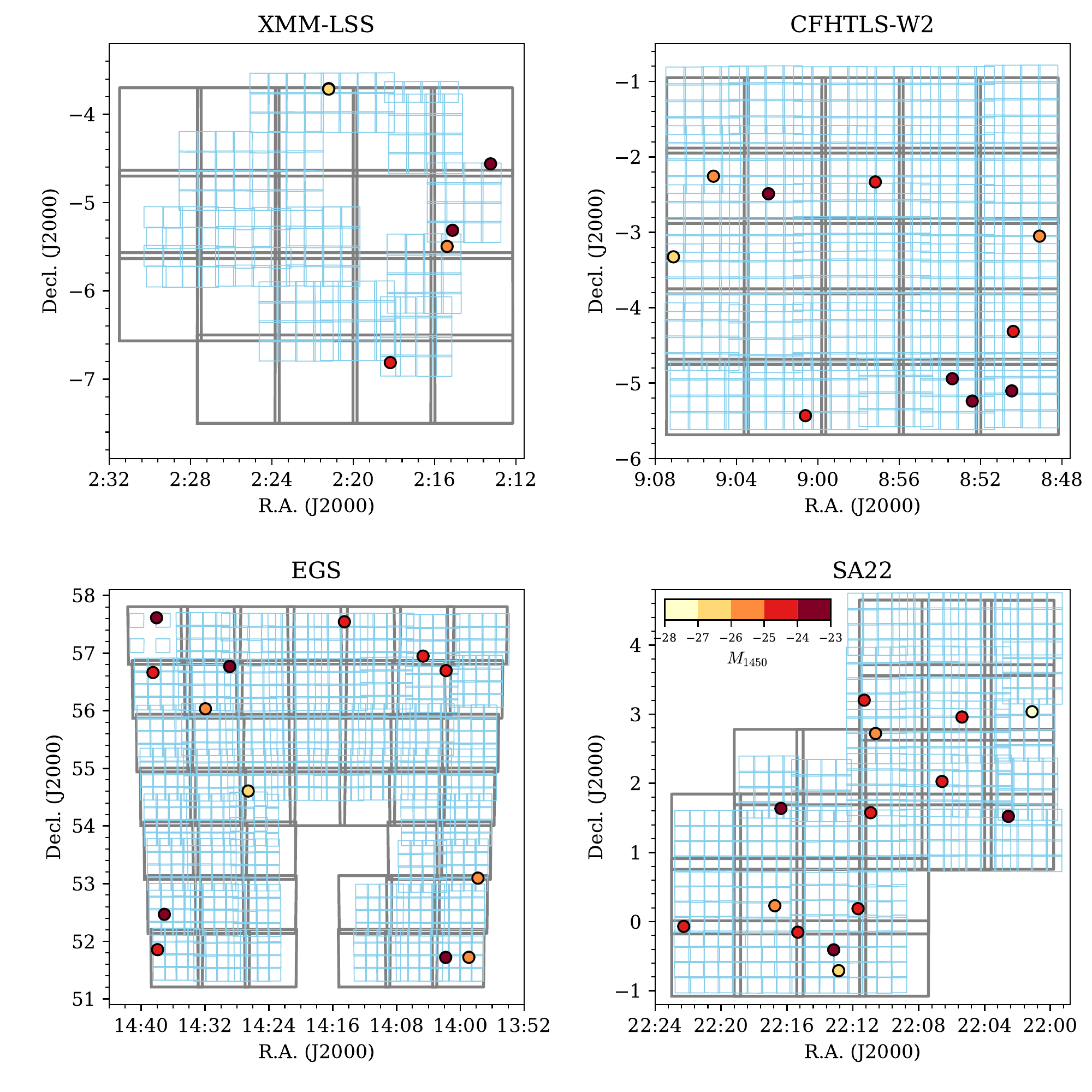}
\caption{
Sky distribution of the IMS quasars. The color map shows the $M_{1450}$ of the quasars. The gray and blue squares represent the tiles of CFHTLS ($1^{\circ}\times1^{\circ}$ for each) and IMS/DXS ($13\farcm65\times13\farcm65$ for each), respectively.
\label{fig:dist}}
\end{figure*}

\begin{deluxetable*}{lllcclllcc}
\tabletypesize{\scriptsize}
\tablecaption{IMS $z\sim5$ Quasar Sample \label{tbl:sample}}
\tablewidth{0pt}
\tablehead{
\colhead{ID} & \colhead{$z$} & \colhead{$M_{1450}$} & \colhead{Case} & \colhead{References} & \colhead{ID} & \colhead{$z$} & \colhead{$M_{1450}$} & \colhead{Case} & \colhead{References}
}
\startdata
IMS J021315$-$043341 & 4.88 & $-23.7$ & 1,2 & (2) & IMS J142635$+$543623 & 4.76 & $-26.3$ & 1,2 & (5) \\
IMS J021507$-$051849 & 4.72 & $-23.3$ & 1 & (1) & IMS J142854$+$564602 & 4.73 & $-24.0$ & 1,2 & (5) \\
IMS J021523$-$052946 & 5.13 & $-25.6$ & 1,2 & (5) & IMS J143156$+$560201 & 4.75 & $-25.3^{a}$ & 1,2 & (5) \\
IMS J021811$-$064843 & 4.87 & $-24.7$ & 1,2 & (2) & IMS J143705$+$522801 & 4.78 & $-23.8^{a}$ & 1,2 & (5) \\
IMS J022112$-$034232 & 4.98 & $-24.3$ & 1,2 & (2) & IMS J143757$+$515115 & 5.17 & $-24.1$ & 1,2 & (5) \\
IMS J022113$-$034252 & 5.02 & $-27.0$ & 1,2 & (5) & IMS J143804$+$573646 & 4.84 & $-23.5$ & 1,2 & (5) \\
IMS J084906$-$030304 & 4.79 & $-25.5$ & 1,2 & (1) & IMS J143831$+$563946 & 4.82 & $-24.5^{a}$ & 1,2 & (5) \\
IMS J085024$-$041850 & 4.80 & $-24.2$ & 1,2 & (2) & IMS J220107$+$030208$^{b}$ & 5.06 & $-27.5$ & 1 & (3) \\
IMS J085028$-$050607 & 5.36 & $-23.5$ & 1 & (2) & IMS J220233$+$013120 & 5.21 & $-23.9$ & 1,2 & (2) \\
IMS J085225$-$051413 & 4.82 & $-23.7$ & 1,2 & (2) & IMS J220522$+$025730 & 4.74 & $-24.4$ & 1 & (2) \\
IMS J085324$-$045626 & 4.83 & $-23.9$ & 1,2 & (2) & IMS J220635$+$020136 & 5.10 & $-24.4$ & 1 & (2) \\
IMS J085711$-$021957 & 4.90 & $-24.1$ & 1 & (1) & IMS J221037$+$024314 & 5.20 & $-25.2$ & 1,2 & (2) \\
IMS J090037$-$052549 & 4.84 & $-25.0$ & 1,2 & (1) & IMS J221055$+$013431 & 4.97 & $-24.6$ & 1,2 & (1) \\
IMS J090226$-$022923 & 4.94 & $-23.7$ & 1 & (1) & IMS J221118$+$031207 & 4.82 & $-24.4$ & 1,2 & (2) \\
IMS J090508$-$021523 & 4.71 & $-25.1$ & 1 & (1) & IMS J221141$+$001119 & 5.23 & $-24.8$ & 1 & (5) \\
IMS J090706$-$031931 & 5.15 & $-26.2$ & 1,2 & (1) & IMS J221252$-$004231 & 4.95 & $-26.3$ & 1,2 & (4) \\
IMS J135747$+$530543 & 5.32 & $-25.5$ & 1,2 & (5) & IMS J221310$-$002428 & 4.80 & $-23.5$ & 1,2 & (5) \\
IMS J135856$+$514317 & 4.97 & $-25.9$ & 1,2 & (5) & IMS J221520$-$000908 & 5.28 & $-24.5$ & 1,2 & (5) \\
IMS J140147$+$564145 & 4.98 & $-24.7$ & 1,2 & (5) & IMS J221622$+$013815 & 4.93 & $-23.3$ & 1,2 & (5) \\
IMS J140150$+$514310 & 5.17$^{c}$ & $-23.4$$^{c}$ & 1 & (5) & IMS J221644$+$001348 & 5.01 & $-25.8$ & 1,2 & (5) \\
IMS J140440$+$565651 & 4.74 & $-24.7^{a}$ & 1,2 & (5) & IMS J222216$-$000406 & 4.95 & $-24.3$ & 1 & (5) \\
IMS J141432$+$573234 & 5.16 & $-24.7$ & 1,2 & (5) &  &  &  &  &  \\
\enddata
\tablecomments{
The ``Case'' columns indicate whether the quasars are included in the case 1 and 2 samples (see details in Section \ref{sec:qsosample}).
The spectral properties are from (1) this work, (2) \citetalias{Kim19}, (3) \cite{Wang16}, (4) \cite{McGreer13}, and (5) \citetalias{McGreer18}.
The differences in cosmological parameters between the literature and this work are also considered.
}
\tablenotetext{a}{These $M_{1450}$ values are not taken from their spectra but from the fitting with their photometric data, as described in \citetalias{Kim19}.}
\tablenotetext{b}{This quasar is excluded in our parametric QLF derivation owing to $M_{1450}<-27$ mag.}
\tablenotetext{c}{These values are revised in \citetalias{Kim19} from those published in \citetalias{McGreer18}.}
\end{deluxetable*}

\subsection{IMS $z\sim5$ Quasar Sample}\label{sec:qsosample}

Including our eight newly discovered quasars, 49 $z\sim5$ quasars with $i'<23$ mag have been spectroscopically identified in our survey area (\citealt{McGreer13}; \citealt{Wang16}; \citetalias{McGreer18,Kim19}).
With these quasars, we constructed the QLF at $z\sim5$.
However, for the QLF construction, we excluded four quasars that were not selected by broadband color (see \citetalias{Kim19}) and two quasars outside the redshift range of $4.7<z<5.4$.
Figure \ref{fig:dist} shows the sky distributions of the remaining 43 quasars, color-coded by their magnitudes.
We used the $z$ and $M_{1450}$ values of eight new quasars measured in Section \ref{sec:specfit}, while those of other quasars were sourced from the literature (\citealt{McGreer13}; \citealt{Wang16}; \citetalias{McGreer18}; \citetalias{Kim19}).

To examine the effect of adding the medium-band color selection, we considered either the broadband selection only (case 1) or a combination of broadband and medium-band selection (case 2).
The numbers for each case are 43 and 32 in cases 1 and 2, respectively, and their $z$ and $M_{1450}$ values are given in Table \ref{tbl:sample}.
The five quasars observed in the three medium bands but outside the selection boxes (the open circles in Figure \ref{fig:ccd}) could have been selected if we obtained deeper images.
However, we excluded them from the case 2 sample and considered this issue in Section \ref{sec:selfun}

\begin{figure}
\centering
\epsscale{1.1}
\plotone{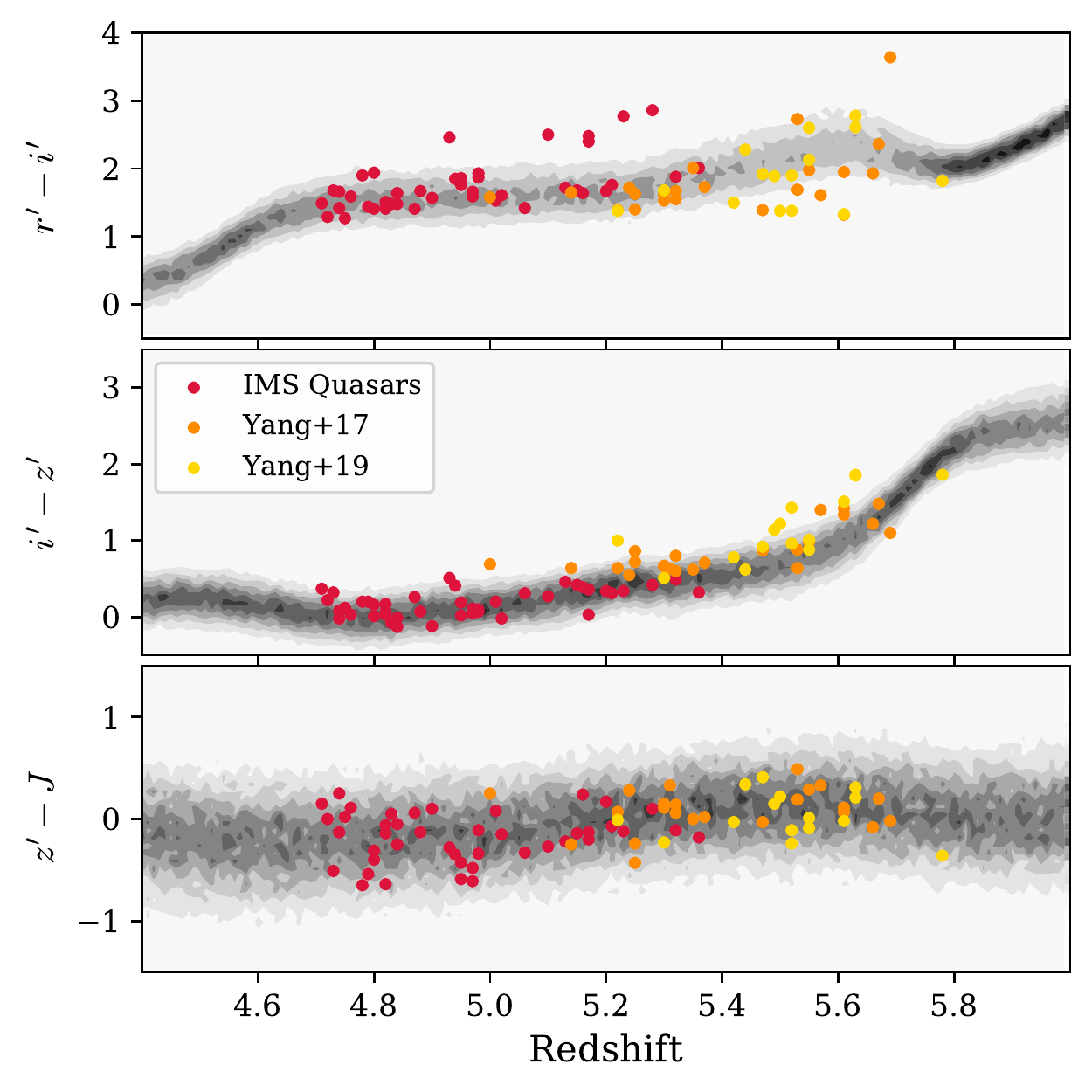}
\caption{
Color distributions of the 100,000 mock quasars (gray contours).
The filled circles are known quasars: IMS quasars in this work (red), \citeauthor{Yang17} (\citeyear{Yang17}; orange), and \citeauthor{Yang19a} (\citeyear{Yang19a}; yellow).
\label{fig:modelcolor}}
\end{figure}

\subsection{Quasar Selection Function}\label{sec:selfun}

\begin{figure*}
\centering
\epsscale{1.}
\plotone{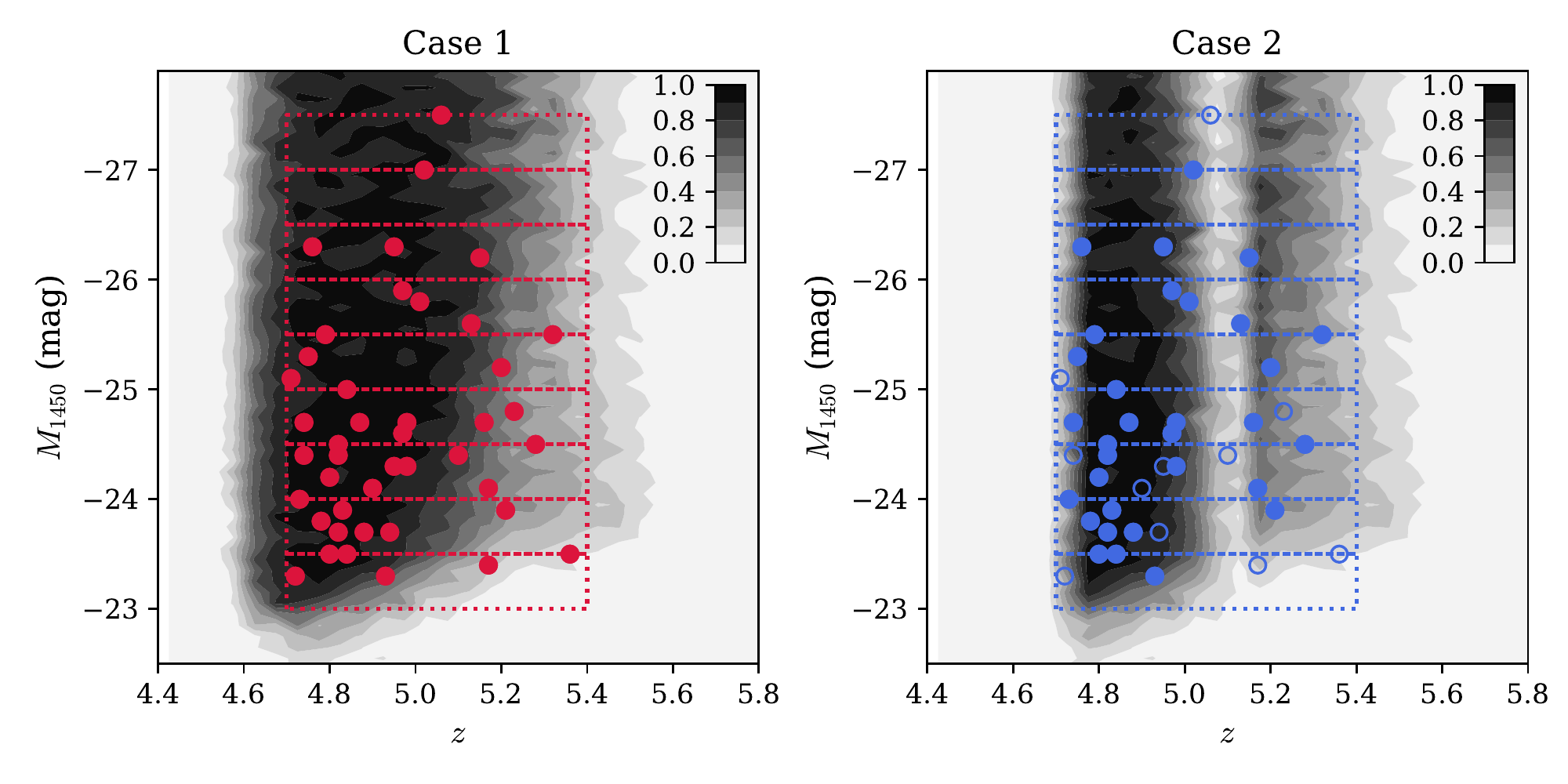}
\caption{
Selection functions for $z\sim5$ quasars in cases 1 (left) and 2 (right).
The contours show the selection efficiencies, for which a scale is given on the inset color bar.
The filled circles denote the spectroscopically identified quasars that are included in each case, while the open circles in the right panel denote those excluded without medium-band observations and/or outside the medium-band selection in case 2.
The bins for deriving our $\Phi_{\rm bin}(M_{1450})$ are shown as boxes with dotted lines.
\label{fig:selfun}}
\end{figure*}

\begin{figure}
\centering
\epsscale{1.1}
\plotone{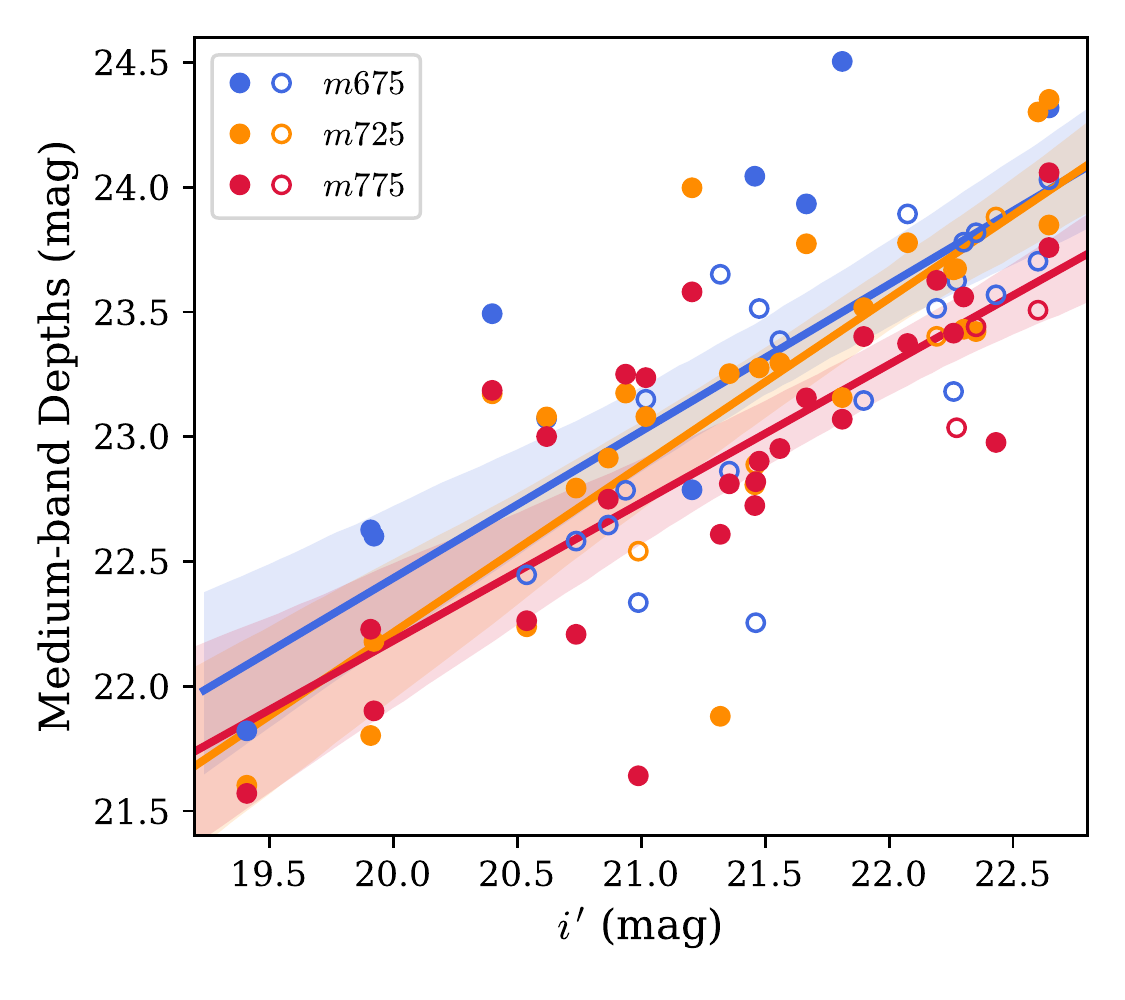}
\caption{
Medium-band imaging depths ($2.7\sigma$) of 32 quasars of case 2 sample along the $i$-band magnitude.
The blue, orange, and red colors denote the cases of $m675$, $m725$, and $m775$, respectively.
The filled/open circles represent whether a quasar is detected/undetected, respectively, while the solid lines with shaded regions are the linear regression models with confidence intervals.
\label{fig:depth}}
\end{figure}

\begin{figure*}
\centering
\epsscale{1.}
\plotone{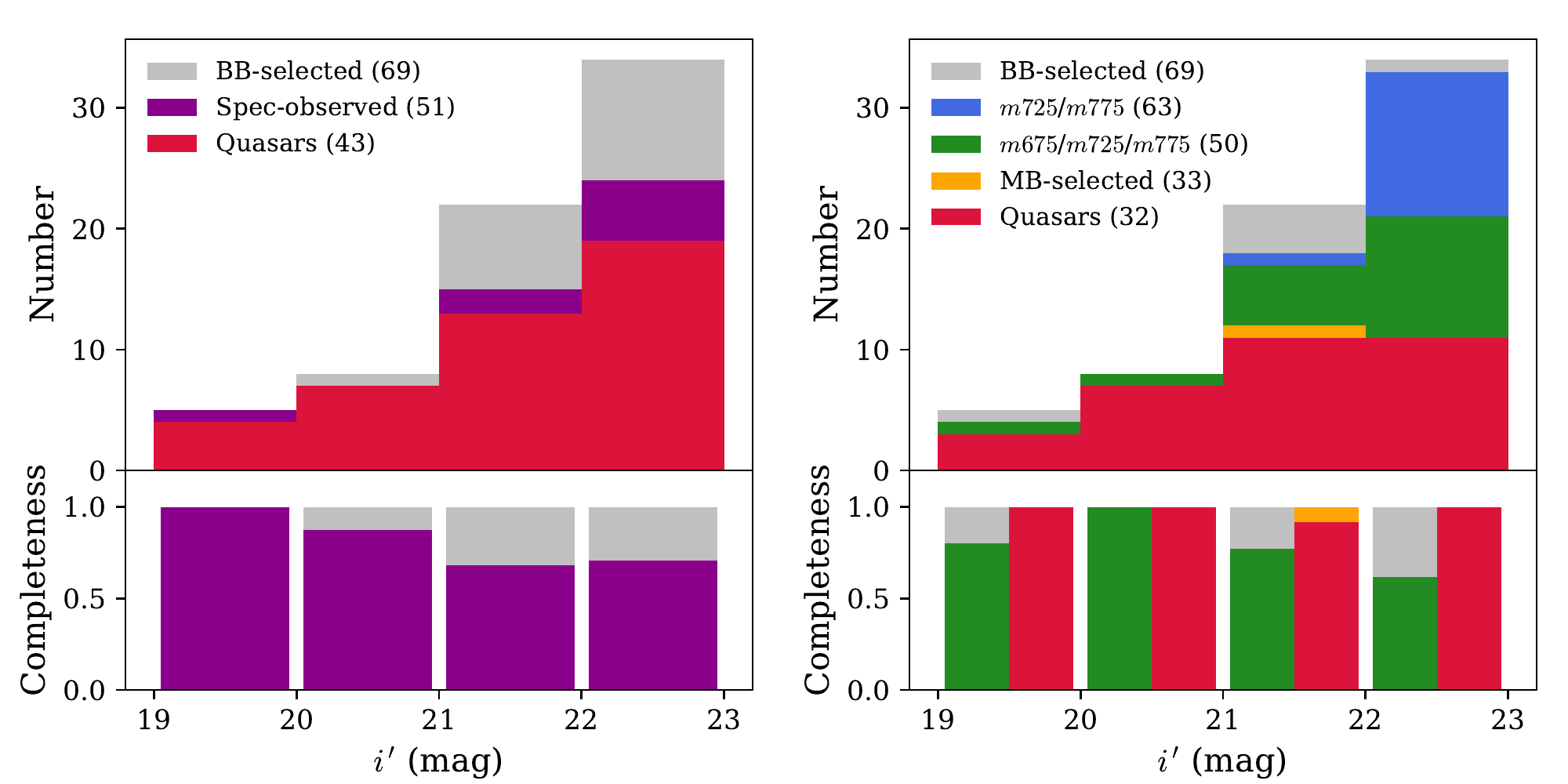}
\caption{
Left: The top panel shows the histogram of candidates along the $i'$-band magnitude of the case 1 sample. 
The gray, purple, and red histograms represent broadband-selected candidates, spectroscopically observed candidates, and quasars identified at $4.7<z<5.4$, respectively.
The bottom panel shows the case 1 spectroscopic completeness.
Right: Histogram (top) and completeness (bottom) of the case 2 sample.
The blue and green histograms show the candidates observed in two ($m725/m775$) and three ($m675/m725/m775$) medium bands, respectively.
The medium-band-selected candidates and the spectroscopically identified quasars at $4.7<z<5.4$ are shown as yellow and red histograms, respectively.
The completenesses shown in the bottom panel are for medium-band photometry (green/gray) and spectroscopy (red/yellow).
\label{fig:hist}}
\end{figure*}

Using our high-redshift quasar model described in Section \ref{sec:specfit}, we calculated the selection efficiency of our color selection criteria.
We generated mock spectra of 100,000 quasars randomly distributed in the range of $4.4<z<6.0$ and $-28<M_{1450}<-22$, while $\alpha_{\lambda}$ and EW$_{\rm Ly\alpha+NV}$ were randomly generated by Gaussian distributions with mean and standard deviation values of $\alpha_{\lambda}=-1.6\pm1.0$ \citep{Mazzucchelli17} and $\log \mathrm{EW}_{\rm Ly\alpha+NV}=1.803\pm0.205$ \citep{Diamond09}, respectively, including the Baldwin effect of Ly$\alpha$ \citep{Dietrich02}.
We integrated the mock quasar spectra over the broadband and medium-band transmission curves to obtain the magnitudes used for color selection.
In Figure \ref{fig:modelcolor}, we show the color distributions of these mock quasars.
They are in line with not only the IMS quasars but also the known $z\sim5.5$ quasars in the same photometric system \citep{Yang17,Yang19a}.
Note that the relatively larger scatter in $r'-i'$ color appears to be due to the various proximity zone sizes of high-redshift quasars according to their ages (e.g. \citealt{Eilers17}).

The imaging survey depths are not always homogeneous, and the varying imaging depths at different locations can affect the selection function.
For case 1, we generated broadband depth maps resampled at a pixel scale of $1\farcm0$.
For each pixel of the broadband depth maps and $z$--$M_{1450}$ bin with a size of $\Delta z =0.05$ and $\Delta M_{1450} = 0.1$ mag, we calculated the selection completeness, a ratio of the number of quasars satisfying our broadband color selection criteria to the total number of mock quasars in the bin (the average number of mock quasars in each bin is $\sim50$).
In this process, we applied additional Gaussian noise to the model magnitudes, considering the image depths and model magnitudes.
In Figure \ref{fig:selfun}, the calculated selection functions are shown as gray contours on the left panel.
Although our $J$-band data were inhomogeneous (\citetalias{Kim19}), the differences between the selection functions of the four extragalactic fields of our survey were negligible.
We used the combined selection function weighted by the area of each survey field.

For case 2, we calculated a selection function including our medium-band color selection criteria.
The imaging depths of the follow-up medium-band observations are not uniform but depend on the $i'$-band magnitudes of the targets, because we gave exposures proportional to the $i'$-band magnitude as much as we could, although the detection limits vary depending on the observing conditions at the time of observation and whether the object was easily detected or not.
If the object was detected with a sufficient S/N ($\gtrsim 10$), we terminated observation of the target in the corresponding band to save the observing time.
Figure \ref{fig:depth} shows the detection limits ($2.7\sigma$) for a point source in the $m675$, $m725$, and $m775$ images of 32 quasars of the case 2 sample, as a function of their $i'$-band magnitude.
We fitted a linear regression model to these imaging depths, shown as the solid lines with confidence intervals calculated with the \texttt{seaborn} Python package\footnote{\url{https://seaborn.pydata.org}}.
Out of 32 quasars, 22 are not detected in $m675$ (open circles), implying small fluxes below the redshifted Ly$\alpha$.
However, we used all of them for linear regression fitting, because we consider such undetected cases in determining the selection function of case 2.

As shown in Figure \ref{fig:ccd}, for objects without detection, we used their detection limits as their magnitudes for the color selection.
In the case of mock quasars, similarly, medium-band magnitudes fainter than the detection limit expected from their $i'$-band magnitudes (i.e., linear regression models) are replaced with the expected values.
The color selection with these modified colors agreed with our medium-band color selection process, which could miss some quasars in marginal conditions (e.g., open circles in Figure \ref{fig:ccd}).
The right panel of Figure \ref{fig:selfun} shows the selection function for case 2.
Note that there is a valley at $z\sim5.1$, which corresponds to a specific redshift for which the medium-band selection fails (see Figure \ref{fig:ccd}).

It has been suggested that the EW$_{\rm Ly\alpha+NV}$ values of $z>5.7$ quasars are lower (e.g., $\log \mathrm{EW}_{\rm Ly\alpha+NV}=1.542\pm0.391$; \citealt{Banados16}) than their low-redshift counterparts, which might be due to the decrease in Ly$\alpha$ transmission at $\lambda>1216\rm{\AA}$ by neutral hydrogen (e.g., \citealt{Davies18}).
At $z\sim5$, the neutral hydrogen fraction is much lower than 0.1 \citep{Fan06,McGreer15}, and thus it has a negligible effect on the redward Ly$\alpha$ fluxes.
If we introduce the EW distribution of \cite{Banados16}, the selection efficiencies of cases 1 and 2 are reduced by about 30\% and 50\% overall, respectively.

\subsection{IMS Survey Completeness}\label{sec:completeness}

We first consider the completeness for case 1.
In the process of selecting the 69 broadband-selected candidates, we used the magnitude cut of $i'<23$ mag, at which the CFHTLS $i'$-band (or $i'_{\rm CFHTLS}$) images we used for the source detection were almost complete (\citealt{Hudelot12}; \citetalias{McGreer18}).
\cite{Hudelot12} provide the 50\% and 80\% completeness limits for a point source of each $i'_{\rm CFHTLS}$-band image.
We fitted these limits with an analytic completeness function \citep{Fleming95}, resulting in $\gtrsim99\%$ completeness at $i'_{\rm CFHTLS}<23$ mag for all CFHTLS images used.
Even if we consider the transformation from $i'_{\rm CFHTLS}$ to $i'$ (i.e., from CFHTLS to SDSS), the point-source detection at $i'<23$ mag is almost complete. 
Therefore, for the broadband photometric completeness, we assume that our detection is complete.
The point-source separation process was not included in our selection\footnote{
\citetalias{McGreer18} adopted $i'_{\rm AUTO}-i'_{\rm PSF} > -0.15$ that showed high completeness of $\sim98\%$ for $i'\lesssim23$ mag sources, where $i'_{\rm{AUTO}}$ is the automatic aperture magnitude (MAG\_AUTO) from SExtractor \citep{Bertin96} and $i'_{\rm{PSF}}$ is the magnitude based on the point spread function (PSF).
From our broadband-selected sample, only three candidates did not satisfy the criteria due to adjacent object(s) or subtle elongation.
Two of them were identified as high-redshift quasars (IMS J022113$-$034252 at $z=5.02$ and IMS J085028$-$050607 at $z=5.36$), while the other was identified as a non-quasar object (IMS J090540$-$011038; see Appendix \ref{sec:addspec}).
The exclusion of these targets does not affect the QLF determination.
}, but note that we performed the visual inspection as mentioned in Section \ref{sec:bbsel}.
The left panel of Figure \ref{fig:hist} shows the number and completeness as a function of the $i'$-band magnitude for case 1.
Of the 69 candidates, 51 were spectroscopically observed and there are 43 quasars at $4.7<z<5.4$ as mentioned above.
We used the ratio of the candidates with spectroscopy to the broadband-selected candidates as the spectroscopic completeness.

For case 2, the completeness for both broadband and medium-band photometry should be considered.
As with case 1, we also assume complete broadband photometry.
In addition, we adopted medium-band photometric completeness.
According to our criteria, quasars at $5.1<z<5.4$ could be selected with the $m725$ and $m775$ bands alone, while 63 of the 69 broadband-selected candidates (91\%) were observed in these two bands.
However, the selection of quasars at $4.7<z<5.1$ requires the $m675$, $m725$, and $m775$ bands, in which 50 of 69 candidates (72\%) were observed.
To avoid complex and ambiguous calculations, the candidates with observations in the three medium bands were considered as the main sample in case 2.
Therefore, we used the ratio of the number of candidates with observations in the three medium bands to the total number of broadband-selected candidates as the medium-band photometry completeness in case 2 (green histogram in the bottom right panel of Figure \ref{fig:hist}).
As described in Section \ref{sec:newquasars}, 33 of the 50 candidates observed in the three medium bands satisfy our medium-band color selection criteria (yellow histogram in Figure \ref{fig:hist}).
We also obtained spectra for 32 of the candidates, all of which were identified as $4.7<z<5.4$ quasars.
The red histogram in the bottom right panel of Figure \ref{fig:hist} shows the spectroscopic completeness of the case 2 sample.

To apply these photometric/spectroscopic completenesses to each case, we generated the photometric/spectroscopic completeness functions of ($z$, $M_{1450}$) using our high-redshift quasar model described in Section \ref{sec:selfun}.
For the mock quasars in each bin of the selection functions, we took the number-weighted mean values of their photometric/spectroscopic completenesses.

\subsection{Binned and Parametric QLFs}\label{sec:binpar}

We first calculated the binned QLF at $z\sim5$ using the $1/V_{a}$ method from \cite{Avni80}.
Given a bin with sizes of $\Delta M_{1450}$ and $\Delta z$, a specific co-moving volume of our survey $V_{a}$ can be calculated as

\begin{figure}
\centering
\epsscale{1.1}
\plotone{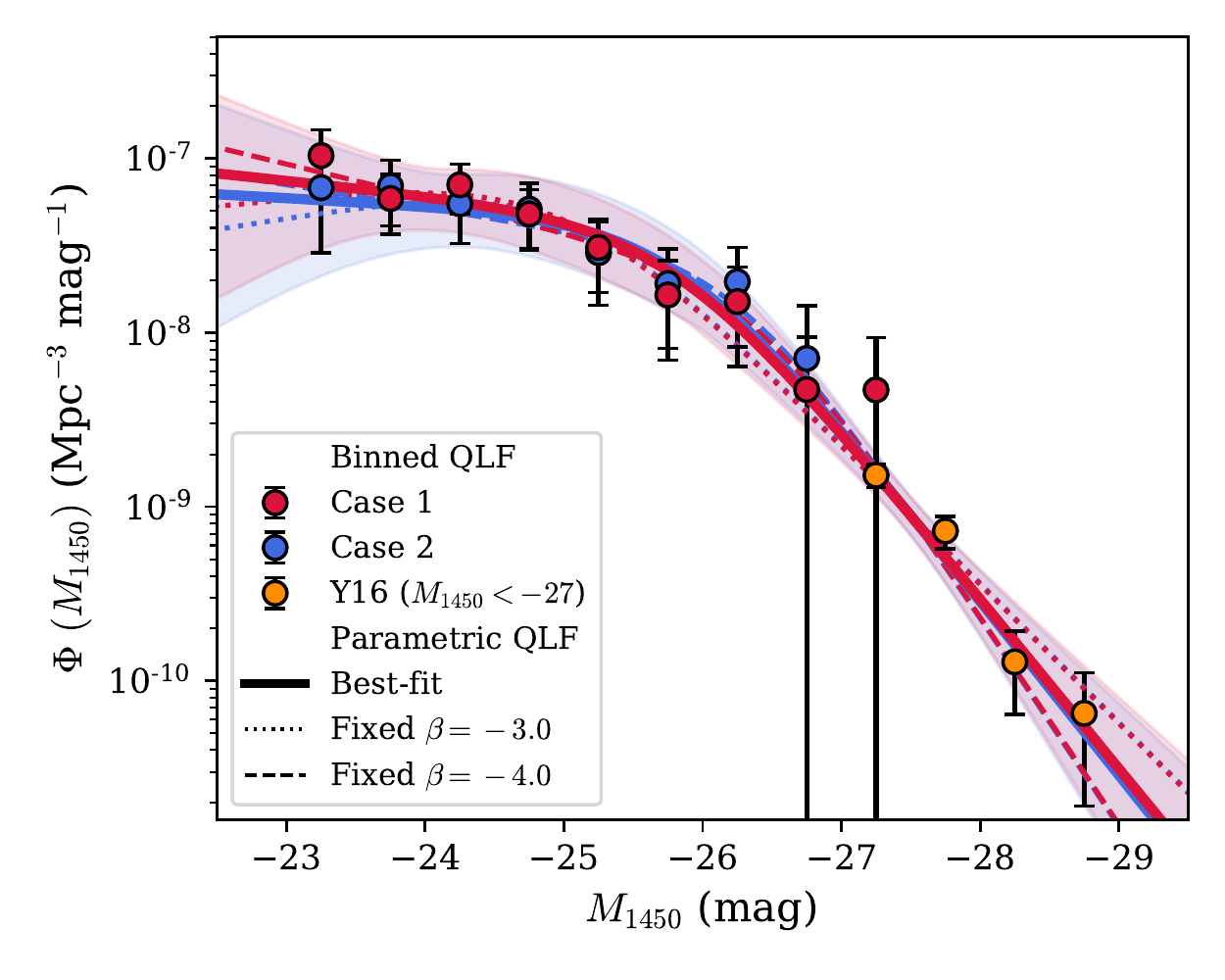}
\caption{
Binned and parametric QLFs of quasars at $4.7<z<5.4$ within the IMS coverage.
The red and blue points are $\Phi_{\rm bin}$ for the IMS $z\sim5$ quasar samples in cases 1 and 2, respectively.
The orange points are that of the \citetalias{Yang16} sample, which are re-binned for bright quasars ($M_{1450}<-27$ mag).
The red and blue solid lines with shaded regions show our best-fit results of $\Phi_{\rm par}$ with 1$\sigma$ confidence level in cases 1 and 2, respectively.
The dotted and dashed lines show the $\Phi_{\rm par}$ with fixed slopes of $\beta=-3.0$ and $-4.0$, respectively, in each case.
\label{fig:qlf}}
\end{figure}

\begin{deluxetable}{ccccccc}
\tabletypesize{\scriptsize}
\tablecaption{Binned QLFs\label{tbl:binqlf}}
\tablewidth{0pt}
\tablehead{
\colhead{} & \multicolumn{2}{c}{Case 1} & \multicolumn{2}{c}{Case 2} & \multicolumn{2}{c}{\citetalias{Yang16} ($M_{1450}<-27$)}\\
\cmidrule(lr){2-3} \cmidrule(lr){4-5} \cmidrule(lr){6-7}
\colhead{$M_{1450}$} & \colhead{$N_{Q}$} & \colhead{$\Phi_{\rm bin}(M_{1450})$} & \colhead{$N_{Q}$} & \colhead{$\Phi_{\rm bin}(M_{1450})$} & \colhead{$N_{Q}$} & \colhead{$\Phi_{\rm bin}(M_{1450})$} 
}
\startdata
$-28.75$ & 0 & ... & 0 & ... & 2 & $0.065\pm0.046$\\
$-28.25$ & 0 & ... & 0 & ... & 4 & $0.128\pm0.064$\\
$-27.75$ & 0 & ... & 0 & ... & 23 & $0.73\pm0.15$ \\
$-27.25$ & 1 & $4.7\pm4.7$ & 0 & ... & 45 & $1.52\pm0.23$\\
$-26.75$ & 1 & $4.7\pm4.7$ & 1 & $7.1\pm7.1$ & 0 & ... \\
$-26.25$ & 3 & $15.1\pm8.7$ & 3 & $19.6\pm11.3$ & 0 & ... \\
$-25.75$ & 3 & $16.5\pm9.5$ & 3 & $19.2\pm11.1$ & 0 & ... \\
$-25.25$ & 5 & $30.8\pm13.8$ & 4 & $28.8\pm14.4$ & 0 & ... \\
$-24.75$ & 7 & $48.1\pm18.2$ & 6 & $51.3\pm20.9$ & 0 & ... \\
$-24.25$ & 10 & $70.7\pm22.4$ & 6 & $55.0\pm22.5$ & 0 & ... \\
$-23.75$ & 7 & $59.0\pm22.3$ & 6 & $69.5\pm28.4$ & 0 & ... \\
$-23.25$ & 6 & $104.1\pm42.5$ & 3 & $68.1\pm39.3$ & 0 & ... \\
\enddata
\tablecomments{We rebinned the \citetalias{Yang16} quasar sample with $M_{1450}<-27$ mag. The size of each magnitude bin is $\Delta M_{1450} = 0.5$ mag.
$\Phi_{\rm bin}$ is in units of 10$^{-9}$ Mpc$^{-3}$ mag$^{-1}$.
}
\end{deluxetable}

\begin{equation}
V_{a}=\frac{1}{\Delta M_{1450}} \int_{\Delta M_{1450}} \int_{\Delta z} p(M_{1450},z) \frac{dV}{dz} dz dM_{1450},
\end{equation}

\noindent where $p(M_{1450},z)$ is the total selection probability combining the quasar selection functions (Section \ref{sec:selfun}) and the photometric/spectroscopic completeness functions (Section \ref{sec:completeness}), and $dV/dz$ is the co-moving volume element of our survey area.
The binned QLF $\Phi_{\rm bin}(M_{1450})$ can then be written as

\begin{equation}
\Phi_{\rm bin}(M_{1450}) = \frac{1}{\Delta M_{1450}} \sum^{N_{Q}} \frac{1}{V_{a}},
\end{equation}

\noindent where $N_{Q}$ is the number of quasars in the bin.
The bin size was set to $\Delta M_{1450}=0.5$ mag (see Figure \ref{fig:selfun}).
After experimenting with several different bin sizes, we settled on this size, which gave a good compromise between the number of available quasars per bin and the fine sampling of the QLF.
Note that the uncertainties of $\Phi_{\rm bin}$ are calculated using the root-sum-square method. 
We have listed the derived $\Phi_{\rm bin}$ for cases 1 and 2 in Table \ref{tbl:binqlf}, which are also denoted by red and blue points in Figure \ref{fig:qlf}, respectively.

We also crosschecked our binned QLF with another nonparametric QLF for IMS quasar sample using the C$^{-}$ estimator \citep{Lynden71}.
The comparison shows that the QLFs derived from different methods agree with each other (see details in Appendix \ref{sec:cminus}).

We also derived the parametric QLF at $z\sim5$ ($\Phi_{\rm par}$), which has a functional form with faint- and bright-end slopes ($\alpha$ and $\beta$, respectively; e.g., \citetalias{McGreer18}):

\begin{equation}
\begin{aligned}
&\Phi_{\rm par}(M_{1450}) = \\ &\frac{\Phi^{*}}{10^{0.4(\alpha+1)(M_{1450}-M_{1450}^{*})}+10^{0.4(\beta+1)(M_{1450}-M_{1450}^{*})}},\label{eq:DPL}
\end{aligned}
\end{equation} 

\begin{figure}
\centering
\epsscale{1.1}
\plotone{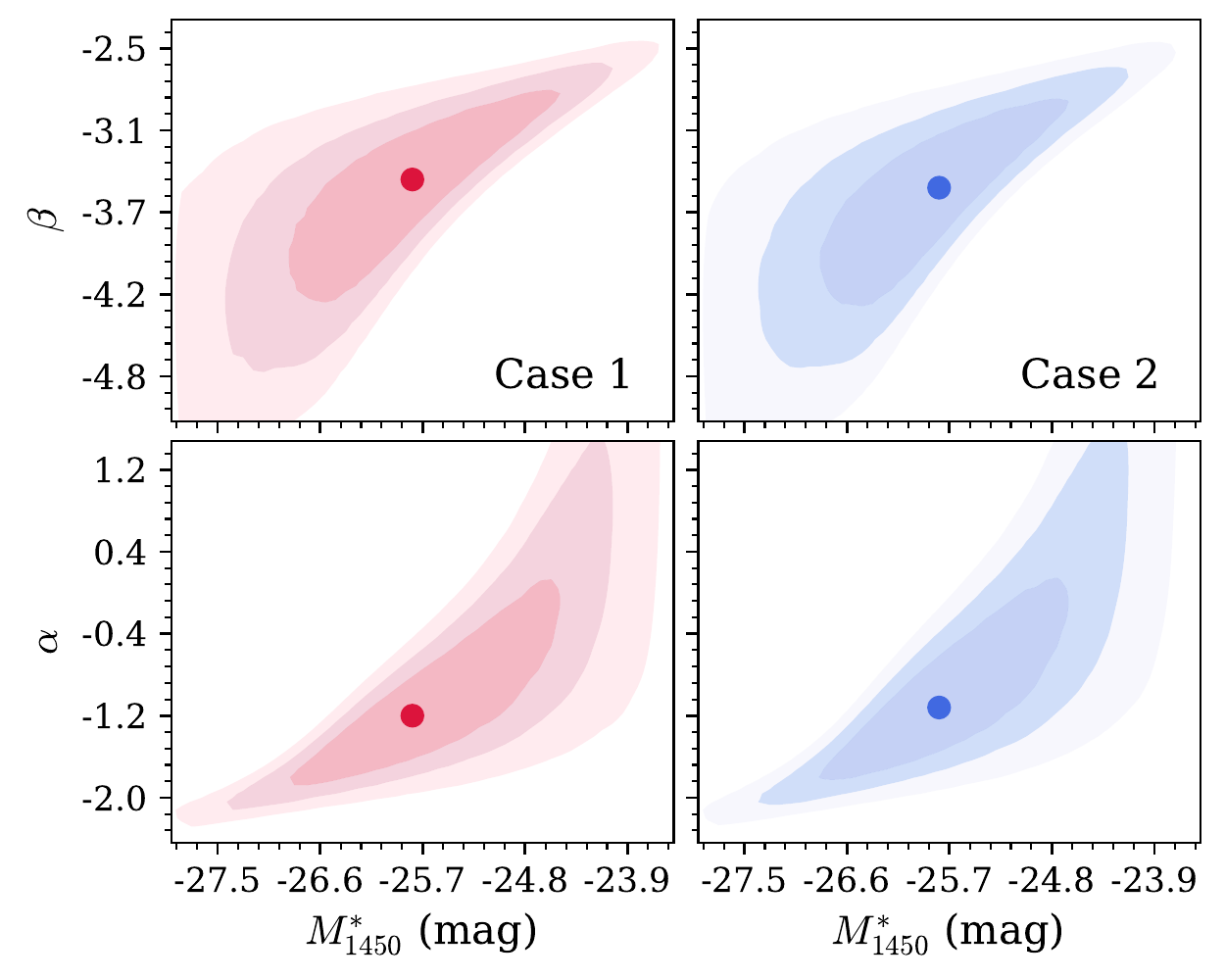}
\caption{
Confidence regions of the best-fit parameters for $\Phi_{\rm par}$ in cases 1 (left) and 2 (right).
The color-filled regions denote the $1\sigma$ (68.3\%), $2\sigma$ (95.4\%), and $3\sigma$ (99.7\%) confidence levels from darker to brighter.
The best-fit values are also marked.
\label{fig:confidence}}
\end{figure}

\begin{deluxetable}{lcccc}
\tabletypesize{\scriptsize}
\tablecaption{Parametric QLFs\label{tbl:parqlf}}
\tablewidth{0pt}
\tablehead{
\colhead{} & \colhead{$\log \Phi^{*}$} & \colhead{$M_{1450}^{*}$} & \colhead{$\alpha$} & \colhead{$\beta$}
}
\startdata
\multicolumn{5}{c}{Case 1}\\
Best-fit & $-7.36_{-0.81}^{+0.56}$ & $-25.78_{-1.10}^{+1.35}$ & $-1.21_{-0.64}^{+1.36}$ & $-3.44_{-0.84}^{+0.66}$ \\
Fixed $\beta$ & $-7.06_{-0.42}^{+0.28}$ & $-25.02_{-0.58}^{+0.39}$ & $-0.79_{-0.70}^{+0.94}$ & $-3.0$ \\
Fixed $\beta$ & $-7.68_{-0.61}^{+0.41}$ & $-26.38_{-0.63}^{+0.39}$ & $-1.48_{-0.43}^{+0.48}$ & $-4.0$ \\
\hline
\multicolumn{5}{c}{Case 2}\\
Best-fit & $-7.35_{-0.77}^{+0.52}$ & $-25.81_{-1.01}^{+1.22}$ & $-1.11_{-0.68}^{+1.31}$ & $-3.50_{-0.81}^{+0.66}$ \\
Fixed $\beta$ & $-7.07_{-0.42}^{+0.30}$ & $-25.04_{-0.57}^{+0.42}$ & $-0.67_{-0.76}^{+1.13}$ & $-3.0$ \\
Fixed $\beta$ & $-7.57_{-0.62}^{+0.39}$ & $-26.28_{-0.63}^{+0.37}$ & $-1.32_{-0.51}^{+0.54}$ & $-4.0$ \\
\enddata
\tablecomments{
$\Phi^{*}$ is in units of Mpc$^{-3}$ mag$^{-1}$.
}
\end{deluxetable}

\noindent where $\Phi^{*}=\Phi^{*}_{z=6}\times10^{k(z-6)}$ with a normalization factor at $z=6$ ($\Phi^{*}_{z=6}$) and $k=-0.47$\footnote{We also tested for $k=-0.7$ calibrated from the bright quasars at $5\lesssim z\lesssim6$ \citep{Jiang16}. There is no significant change in the resultant parameters within the 1$\sigma$ errors at this narrow redshift range.\label{foot:k}} \citep{Fan01b},  and $M_{1450}^{*}$ is the break magnitude.
Accounting for the lack of bright quasars ($M_{1450}<-27$ mag) in our sample, we used an ancillary SDSS quasar sample (\citetalias{Yang16}).
The IMS and \citetalias{Yang16} samples overlap at $M_{1450}\sim-27$ mag.
We simply assigned the two samples to cover different ranges; we used the IMS sample at $M_{1450}\geq-27$ mag and the \citetalias{Yang16} sample at $M_{1450}<-27$ mag.
This is because of the potentially underestimated number of \citetalias{Yang16} quasars at $M_{1450}\gtrsim-27$ mag, which is lower than that of \citetalias{McGreer18} by a factor of 2--5.
For ease of comparison, we re-binned the QLF of the \citetalias{Yang16} sample ($4.7<z<5.4$ and $M_{1450}<-27$ mag) and marked them with orange points in Figure \ref{fig:qlf}.
Note that we used the completeness functions for the SDSS quasars, provided by Jinyi Yang personally.

We used the maximum likelihood method including selection efficiencies \citep{Marshall83} to fit the function with four parameters ($\Phi^{*}$, $M_{1450}^{*}$, $\alpha$, $\beta$) by minimizing $S$, defined as

\begin{equation}
\begin{aligned}
S =& -2 \sum \ln \left[ \Phi_{\rm par}(M_{1450},z) p(M_{1450},z) \right] \\
&+ 2 \iint \Phi_{\rm par}(M_{1450},z) p(M_{1450},z) \frac{dV}{dz} dz dM_{1450}.
\end{aligned}
\end{equation}

\noindent The first term of this equation is the sum of all the IMS and \citetalias{Yang16} samples. 
The second term stands for the expected number of quasars with given $\Phi_{\rm par}$ and $p$, which is integrated over the ranges of $-29.0 < M_{1450} < -23.0$ and $4.7<z<5.4$.
Therefore, minimizing $S$ is equivalent to finding the appropriate $\Phi_{\rm par}$ to satisfy both the observations and expectations.
In Figure \ref{fig:qlf}, the best-fit results of $\Phi_{\rm par}$ of cases 1 and 2 are shown as red and blue solid lines, respectively, in line with the $\Phi_{\rm bin}$ of each case.
Note that the uncertainties of the free parameters are calibrated from the $\Delta S = S - S_{\rm min}$ distribution, under the assumption of $\Delta S \sim \chi^{2}_{\nu}$ \citep{Lampton76}, as shown in Figure \ref{fig:confidence}.
The resultant $\Phi_{\rm par}$ parameter of case 1 (case 2) has a break magnitude of $M_{1450}^{*}=-25.78^{+1.35}_{-1.10}$ mag ($-25.81^{+1.22}_{-1.01}$ mag) with a faint-end slope of $\alpha=-1.21^{+1.36}_{-0.64}$ ($-1.11^{+1.31}_{-0.68}$) and a bright-end slope of $\beta=-3.44^{+0.66}_{-0.84}$ ($-3.50^{+0.66}_{-0.81}$).
Interestingly, the derived slopes and break magnitudes of our QLF are similar to those of the QLFs at adjacent redshifts of $z=4$ and $6$ (\citealt{Akiyama18,Matsuoka18b}), and this will be discussed in a forthcoming paper (Y. Kim et al. 2020, in preparation).
We also obtained the $\Phi_{\rm par}$ parameters with fixed bright-end slopes of $\beta=-3.0$ and $-4.0$ (the dotted and dashed lines in Figure \ref{fig:qlf}, respectively).
These are consistent with the best-fit results within the $1\sigma$ confidence level.
All of the derived values are listed in Table \ref{tbl:parqlf}.

When we compare the estimated QLFs of cases 1 and 2, they are consistent with each other, in both the $\Phi_{\rm bin}$ and $\Phi_{\rm par}$ cases.
The only minor differences between them are observed at $M_{1450}>-25$ mag, especially at the faintest bin.
This possible difference at the faintest bin is due to the serendipitous identification of faint quasars, such as IMS J085028$-$050607 with $M_{1450}=-23.5$ mag at $z=5.36$.
As shown in Figure \ref{fig:selfun}, the quasar is located at a very low completeness region and the $V_{a}$ of the faintest bin is smaller than the other bins, giving a high number density at the faintest bin after the completeness correction.
In fact, in case 2 where this quasar is not included, the $\Phi_{\rm bin}(-23.25)$ is more in line with our best-fit QLF. 
But here, we stress that the faintest bin is also consistent with our best-fit QLF within 1$\sigma$ confidence level.

Considering the high success rate of our medium-band selection, especially at fainter magnitudes (\citetalias{Kim19}), a future wide-field imaging survey in medium bands will allow us to estimate reliable QLFs at high redshifts even without deep spectroscopic observations.
In the following discussion, the best-fit result in case 1 is used as a representative result.

\begin{figure}
\centering
\epsscale{1.2}
\plotone{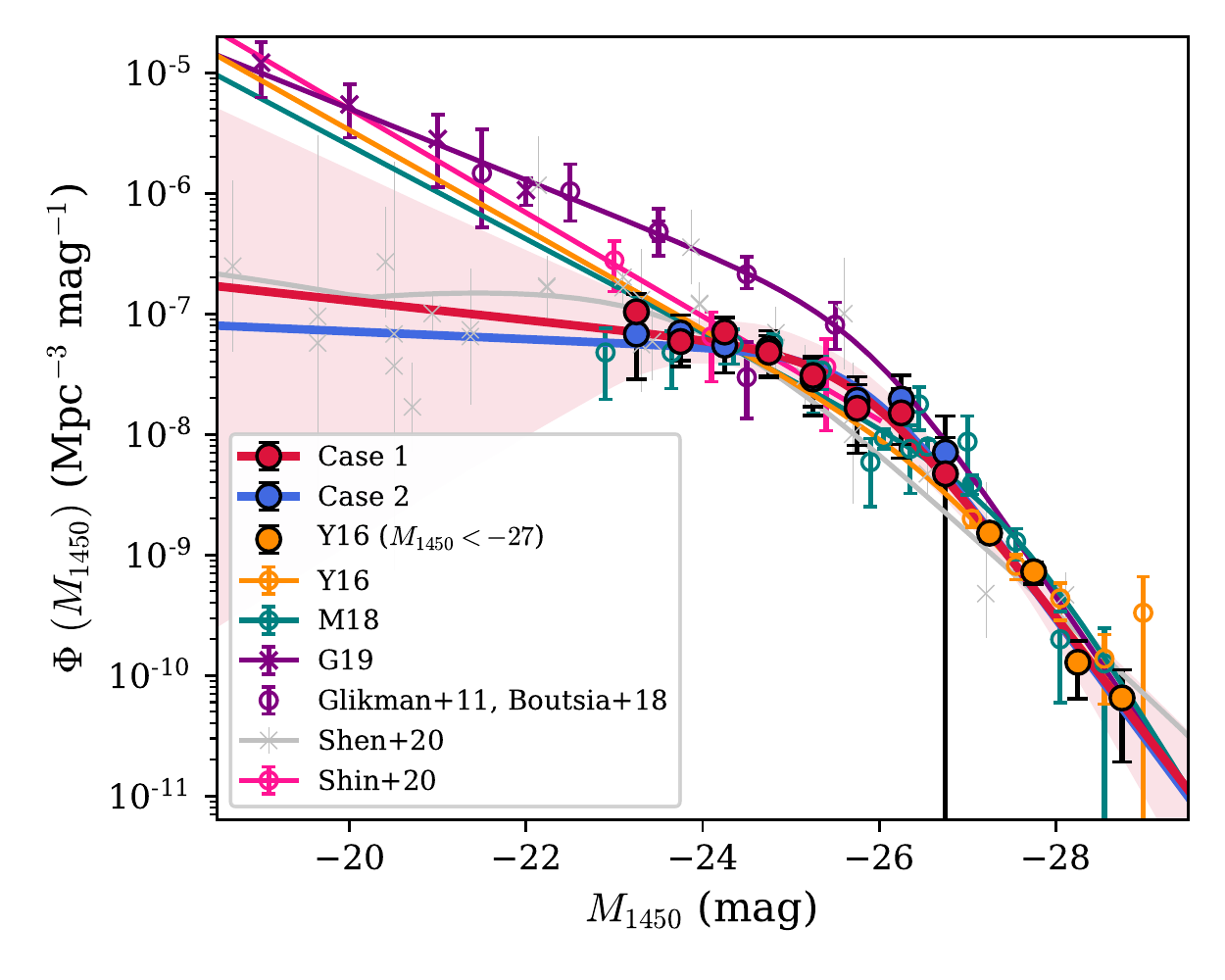}
\caption{
QLFs at $z\sim5$.
The red filled circles show our binned QLF in case 1, while the red solid line with a shaded region indicates the best-fit parametric QLF with 1$\sigma$ confidence level.
The blue filled circles with the blue solid line are those in case 2.
The orange filled circles show the re-binned \citetalias{Yang16} QLF consisting of the bright quasars that we used for fitting.
The open circles and solid lines are the QLFs from the optical/NIR surveys: \citetalias{Yang16} (orange), \citetalias{McGreer18} (teal), and \citeauthor{Shin20} (\citeyear{Shin20}; pink).
The X-ray QLFs of \citetalias{Giallongo19} (purple) and \citeauthor{Shen20} (\citeyear{Shen20}; gray) are shown as crosses with solid lines, which are shifted to $z=5$ (see text).
The QLFs of \cite{Glikman11} and \cite{Boutsia18} from the optical/NIR surveys, which are used by \citetalias{Giallongo19}, are represented by purple open circles for ease of comparison.
In the case of \cite{Shen20}, we only plotted their X-ray QLFs, while their global QLF (solid line) was derived from not only X-ray but also UV/optical quasars.
\label{fig:z5qlfs}}
\end{figure}

For the IMS quasars only, we also performed testing with a single power-law function: $\Phi_{\rm par} \propto 10^{-0.4(\alpha+1)}$.
The resultant slopes are $\alpha=-1.70^{+0.24}_{-0.26}$ and $-1.67^{+0.29}_{-0.31}$ in cases 1 and 2, respectively, which are slightly steeper than our best-fit results.
These values agree with the recent estimate of $\alpha=-2.1\pm0.7$, obtained from a smaller sample of $z\sim5$ faint quasars using a single power-law function (pink line in Figure \ref{fig:z5qlfs}; \citealt{Shin20}).

\section{DISCUSSIONS}\label{sec:discussion}

\subsection{Reliability of the $\lowercase{z}\sim5$ QLF\label{sec:compare}}

Figure \ref{fig:z5qlfs} shows various QLFs at $z\sim5$ from the literature (\citealt{Glikman11}; \citetalias{Yang16}; \citealt{Boutsia18}; \citetalias{McGreer18}; \citetalias{Giallongo19}; \citealt{Shen20,Shin20}), in comparison to our best-fit QLFs.
To obtain the binned QLF of \citetalias{Giallongo19}, we averaged their binned QLFs at $z=4.5$ and $5.6$ (purple crosses). Note that the upper and lower limits of each point indicate the binned QLFs at $z=4.5$ and 5.6, respectively.
Their parametric QLF (model 2; purple solid line) is also adjusted from $z=4.5$ to $z=5$ with $k=-0.47$.
Considering that the binned QLFs of \cite{Glikman11} and \cite{Boutsia18} are used for the parametric QLF derivation of \citetalias{Giallongo19}, for ease of comparison, they are represented by purple open circles.
Similarly, the $z\sim4.8$ QLF of \citeauthor{Shen20} (\citeyear{Shen20}; gray crosses) is also shifted to $z=5$.

While the faint-end slope of our parametric QLF has a flatter value, the binned QLFs in this work are consistent with most QLFs at $-25<M_{1450}<-23$ except for the result of \citetalias{Giallongo19} and results from their group in previous years.
Possible reasons for the discrepancy have been discussed in detail in \cite{Parsa18}, but here we discuss if the discrepancy is caused by cosmic variance.
In this magnitude range, \citetalias{Giallongo19}  used the $z\sim4$ QLFs (open purple circles) determined from small-area surveys (3.76 deg$^{2}$ of \citealt{Glikman11}; 1.73 deg$^{2}$ of \citealt{Boutsia18}).
We randomly selected 1.73 deg$^{2}$ area in a circular shape from our survey area, and counted the numbers of $M_{1450}=-23.5$ and $-24.5$ mag quasars ($\Delta M_{1450}=1$ mag), which is repeated 100,000 times.
Then, we examined the frequency of obtaining the $\Phi_{\rm bin}$ values of \cite{Boutsia18} or higher at the two magnitude bins by chance due to cosmic variance; only 2.2\% and 1.4\% for $M_{1450}=-24.5$ and $-23.5$ mag, respectively.
If we require the number densities to be high in these two consecutive magnitude bins, then the probability drops to zero, which can be also inferred from Figure \ref{fig:dist}.
We also tested for 5.49 deg$^{2}$ area, resulting in the very low probability of $\ll1\%$ even if only one magnitude bin was considered.
The result shows that the likelihood of gaining a high quasar number density due to cosmic variance is small, even with an area of $1.73$ deg$^{2}$.

All quasars that we identified with spectroscopy (\citetalias{Kim19} and this work) have EW$_{\rm Ly\alpha+NV}$ higher than 15.4 $\rm\AA$, the criterion for weak Ly$\alpha$ quasars \citep{Diamond09}.
The fraction of weak Ly$\alpha$ quasars is known to be 13.7\% at $z>5.7$ \citep{Banados16}, higher than the fraction of 1.8\% from the EW distribution adopted for our calculation of the selection function \citep{Diamond09}.
Using the EW distribution of \cite{Banados16} for our calculation of the selection function instead (see Section \ref{sec:selfun}), we find that our QLF could be underestimated by a factor of $\sim1.3$ at the faint end (both in cases 1 and 2).
The exact value of this factor is more likely to be smaller than 1.3, assuming the gradual change in the EW distributions between $z\sim5$ and 6.
Future investigation of the EW distributions is desired to improve the QLF estimate.

The \citetalias{Giallongo19} QLF at $M_{1450}>-23$ mag mainly relies on X-ray-selected quasars that do not always overlap with those selected in the UV/optical range.
One possible explanation for their higher number density than UV/optical quasars is due to the high fraction of obscured X-ray quasars at high redshifts (e.g., 50--80\% at $3<z<6$; \citealt{Vito18}), if the obscured fraction applies equally to UV/optical selected quasars.
However, recent X-ray QLFs provided by \citeauthor{Shen20} (\citeyear{Shen20}; gray crosses), corrected by their model prediction, also agree with our results, reinforcing the suggestion of the overestimation of \citetalias{Giallongo19} even at $M_{1450}>-22$ mag.
We note that the $\Phi_{\rm bin}$ of \citetalias{McGreer18} has a value lower than those of our and other QLFs at the faintest bin they explored ($M_{1450} > -23$ mag).
QLFs can be over- or underestimated easily at the faintest end depending on how well the selection function is constructed.
We suggest this as a possible reason for the discrepancy.

\begin{figure}
\centering
\epsscale{1.2}
\plotone{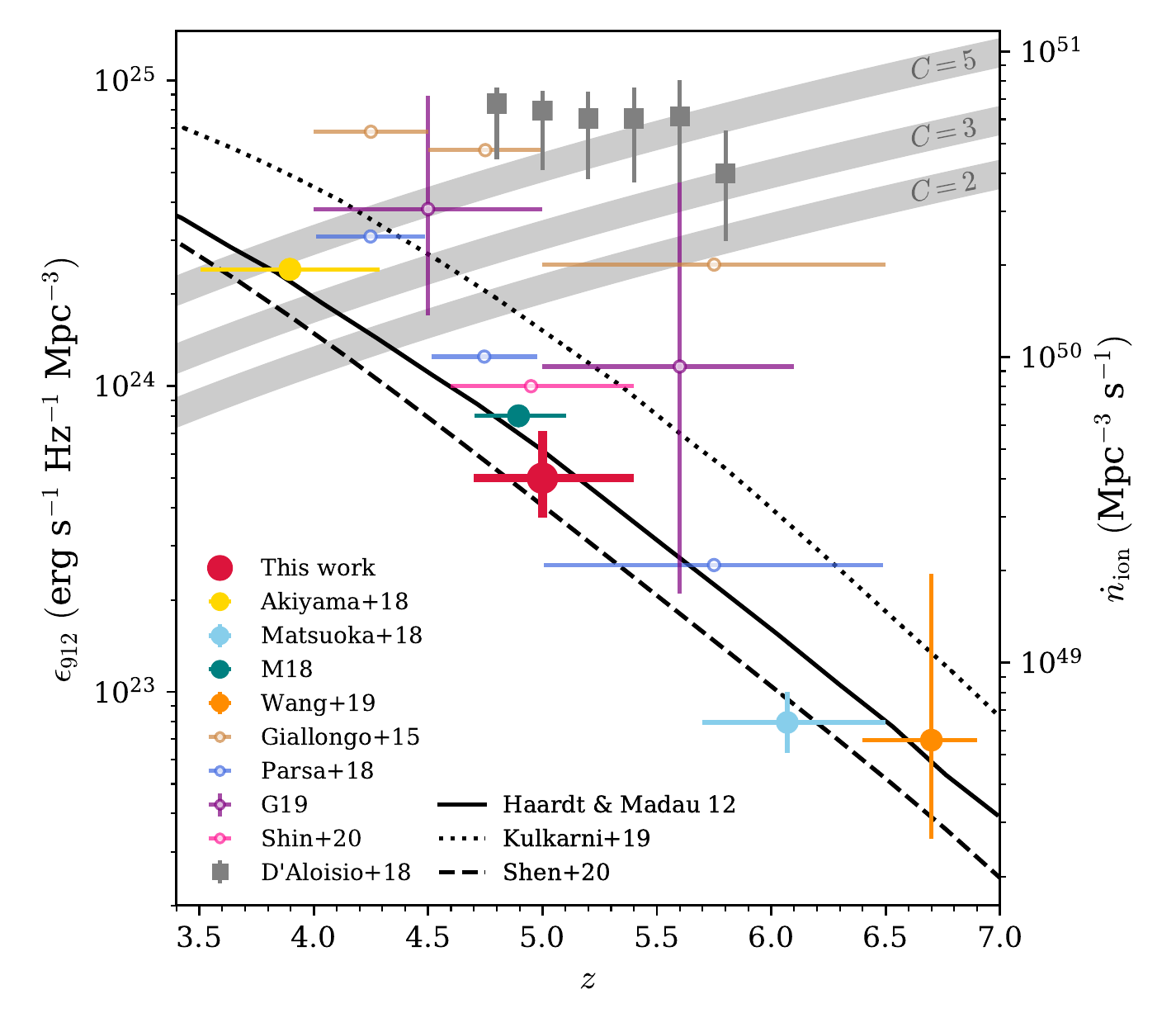}
\caption{
Observed $\epsilon_{912}$ of high-redshift quasars integrated down to $M_{1450}<-18$ mag: this work (red), \citeauthor{Giallongo15} (\citeyear{Giallongo15}; brown), \citeauthor{Akiyama18} (\citeyear{Akiyama18}; yellow), \citeauthor{Matsuoka18b} (\citeyear{Matsuoka18b}; skyblue), \citetalias{McGreer18} (teal), \citeauthor{Parsa18} (\citeyear{Parsa18}; blue), \citetalias{Giallongo19} (purple), and \citeauthor{Wang19} (\citeyear{Wang19}; orange).
The values determined from the small-area surveys ($<10$ deg$^{2}$) are shown as open circles.
The horizontal line on each point indicates the redshift range of the sample used in each study, and the vertical line indicates 1$\sigma$ error.
The solid, dotted, and dashed lines are the emissivity curves from \cite{Haardt12}, \cite{Kulkarni19}, and \cite{Shen20}, respectively.
The thick solid gray lines show the emissivity values required to fully ionize the IGM as a function of redshift, for a given clumping factor $C~(=$2, 3, 5) and $f_{\rm esc}=1$ \citep{Madau99}.
The right-hand vertical axis shows the corresponding $\dot{n}_{\rm ion}$ for a given $\epsilon_{912}$, assuming the UV spectral shape of \cite{Lusso15} and $f_{\rm esc}=1$.
The gray squares show the photon density required to maintain the ionization state of IGM, predicted from the transmitted Ly$\alpha$ flux measurements \citep{Daloisio18}.
\label{fig:eps}}
\end{figure}

\subsection{Contribution of $\lowercase{z}\sim5$ Quasars to Ionizing Background}

We followed the method in \cite{Kulkarni19} to calculate the UV ionizing emissivity of $z\sim5$ quasars at 1450 ($\epsilon_{1450}$) and 912 $\rm\AA$ ($\epsilon_{912}$), which can be written as

\begin{equation}
\epsilon_{1450} = \int_{-30}^{-18} \Phi_{\rm par}(M_{1450})~10^{-0.4(M_{1450}-51.60)}~dM_{1450},\label{eq:emi}
\end{equation}

\noindent and

\begin{equation}
\epsilon_{912} = \epsilon_{1450}\times \left(\frac{912}{1450}\right)^{0.61}\label{equ:e912}.
\end{equation}

\noindent To evaluate $\epsilon_{912}$ from $\epsilon_{1450}$, we assume a double power-law shape of the quasar UV spectra ($f_{\nu} \propto \nu^{-\alpha_{\nu}}$, where $\alpha_{\nu}=0.61$ if $\lambda \geq 912~\rm\AA$ and $\alpha_{\nu}=1.70$ if $\lambda<912~\rm\AA$; \citealt{Lusso15}).
For the best-fit case 1 QLF, $\epsilon_{1450}$ and $\epsilon_{912}$ are $6.64^{+2.81}_{-1.70}$ (or 4.94 to 9.45) and $5.00^{+2.12}_{-1.28}$ (or 3.72 to 7.12), respectively, in units of $10^{23}$ erg s$^{-1}$ Hz$^{-1}$ Mpc$^{-3}$.
Figure \ref{fig:eps} shows the observed $\epsilon_{912}$ values of high-redshift quasars.
Our result is the lowest value at $z\sim5$, an order of magnitude lower than \cite{Giallongo15} estimates (brown open symbols).
Considering large errors, most of the $\epsilon_{912}$ values are consistent with the synthesis model of \cite{Haardt12} and the best-fit model of \cite{Shen20}.
Notable exceptions to the above trend are the results of \cite{Giallongo15} and \citetalias{Giallongo19}.
However, the former were derived from X-ray candidates without spectroscopic confirmation, while the latter are also less reliable than others due to the possible overestimation of their QLF as mentioned earlier in this section.

The above results, however, could be sensitive to the uncertain extrapolation of the QLF down to faint magnitudes of $M_{1450}=-23$ to $-18$ mag.
If we integrate Equation (\ref{eq:emi}) up to $M_{1450}=-23$ mag, at which our QLF is well constrained, $\epsilon_{912}$ is $4.48\times 10^{23}$ erg s$^{-1}$ Hz$^{-1}$ Mpc$^{-3}$.
This is consistent with $\epsilon_{912}$ integrated up to $M_{1450}=-18$ mag within 1$\sigma$ error, meaning that the additional contribution of quasars with $M_{1450}>-23$ mag is not significant unless their number density is unexpectedly high.

Despite the uncertainty in our emissivity estimates at the faintest end, we can explore the significance of the contribution of quasars to the UV ionizing radiation.
Under the same assumptions, we calculate the number density of ionizing photons,

\begin{equation}
\dot{n}_{\rm ion}=f_{\rm esc} \epsilon_{1450} \xi_{\rm ion},
\end{equation}

\noindent where $f_{\rm esc}$ is an escape fraction assumed to be unity ($f_{\rm esc}=1$; e.g., \citealt{Grazian18}), and $\xi_{\rm ion}$ is the ionizing photon production efficiency of a quasar, normalized at 1450 $\rm \AA$.
We obtain $\xi_{\rm ion}\simeq 6.05 \times 10^{25}$ Hz erg$^{-1}$ with the quasar spectral shape of \cite{Lusso15}.
Using the $\epsilon_{1450}$ value from our $z\sim5$ QLF, we obtain $\dot{n}_{\rm ion}$ of $(4.02^{+1.70}_{-1.03})\times10^{49}$ Mpc$^{-3}$ s$^{-1}$.
At $z=5$, the required (critical) photon density to keep balance with hydrogen recombination is $\dot{n}^{\rm cri}_{\rm ion}\sim6.3\times10^{49}~C$ Mpc$^{-3}$ s$^{-1}$ (\citealt{Madau99,Bolton07}; thick gray lines in Figure \ref{fig:eps}), where $C$ is the \ion{H}{2} clumping factor recently predicted as $C<5$ at $z=5$ \citep{Bolton07,Shull12,Jeeson14,Daloisio20}.
For a plausible clumping factor of $C=3$, our result suggests that $z\sim5$ quasars radiate about 21$^{+9}_{-5}$\% of the UV ionizing photons required to balance the ionized state of hydrogen at that time.
This fraction can reach $45$\% if $C=2$ and $\dot{n}_{\rm ion}$ is the 1$\sigma$ upper value of $5.72\times10^{49}$ Mpc$^{-3}$ s$^{-1}$.
In contrast, if we adopt the 1$\sigma$ lower limit, the quasar contribution is $9$\%.
Therefore,  $z\sim5$ quasars alone radiate only up to $\sim50\%$ of the minimum requirement of UV photons to balance with hydrogen recombination rates ($\dot{n}^{\rm cri}_{\rm ion}$ of \citealt{Madau99}), under the extreme assumptions given above.

An alternative method to estimate $\dot{n}^{\rm cri}_{\rm ion}$ is by using observational inference, such as the Ly$\alpha$ forest.
Previous studies calculated the photoionization rate of the UV ionizing background required to match the mean transmitted Ly$\alpha$ flux of high-redshift quasars \citep{Bolton07,Wyithe11,Daloisio18}, and the corresponding critical emissivity $\epsilon_{912}^{\rm cri}$.
The $\dot{n}^{\rm cri}_{\rm ion}$ can then be approximated to $\dot{n}^{\rm cri}_{\rm ion} \simeq (\epsilon_{912}^{\rm cri}/h_{p}\alpha_{\nu})$ Mpc$^{-3}$ s$^{-1}$ \citep{Meiksin05,Becker13}, where $h_{p}$ is the Planck constant and $\alpha_{\nu}=-1.70$ \citep{Lusso15}.
In Figure \ref{fig:eps}, the gray squares indicate the predicted $\dot{n}^{\rm cri}_{\rm ion}$ values with the recently estimated $\epsilon_{912}^{\rm cri}$ values of \cite{Daloisio18}.
At $z=5$, we obtain $\dot{n}^{\rm cri}_{\rm ion}=6.4^{+1.1}_{-2.3}\times10^{50}$ erg s$^{-1}$, indicating that the quasar contribution with our QLF is only approximately $6^{+8}_{-2}\%$ (or 4--14\%) considering the $1\sigma$ errors of $\dot{n}_{\rm ion}$ and $\dot{n}^{\rm cri}_{\rm ion}$.
Such a low contribution is similar to that at $z\sim6$, where the quasar contribution is considered to be less than 10\% \citep{Willott10b,Kashikawa15,Kim15,Onoue17,Matsuoka18b}.
However, note that the $\epsilon_{912}^{\rm cri}$ determination from the Ly$\alpha$ opacity of high-redshift quasars is sensitive to the mean free path of hydrogen photons, which can also be affected by the main ionizing sources \citep{Daloisio18}.

Unlike at $z\sim6$, there is a discrepancy between the $\dot{n}_{\rm ion}^{\rm cri}$ from the two methods at $z=5$.
This is due to the increases in the number of ionizing sources and the volume filled with ionized hydrogen from $z=6$ to 5.
Considering this difference, we conclude that quasars are likely to occupy only a minor portion of the total UV ionizing background at $z=5$, estimated from the Ly$\alpha$ opacity of high-redshift quasars.
However, they can provide nearly half the quantity of photons required to balance with recombination rates.
A better understanding of the quasar contribution for $z=5$ IGM ionizing photons requires the construction of a QLF at $M_{1450}>-23$ mag where we rely on extrapolation to estimate the QLF shape.

In addition, improving constraints on $C$ and other physical properties of high-redshift quasars (e.g., $f_{\rm esc}$) is also required.
For instance, our calculation of $\dot{n}_{\rm ion}/\dot{n}_{\rm ion}^{\rm cri}$ depends on the assumed UV spectral slope of quasars.
The assumed value of $\alpha_{\lambda}=\alpha_{\nu}-2=-1.39$ \citep{Lusso15} is based on quasars at $z=2.4$ and may not represent UV spectral slopes of $z=5$ quasars in this work.
For example, \cite{Mazzucchelli17} find $\alpha_{\lambda}=-1.6\pm0.1$ for $\bar{z}=6.6$ quasars, which is steeper than the \cite{Lusso15} value.
We directly infer the UV slope of our sample using the SED-fitting result of a sub-sample of our $z=5$ quasars in \citetalias{Kim19}. 
Using the values presented in Table 5 of \citetalias{Kim19}, we find $\alpha_{\lambda} = -1.9 \pm 0.1$, which is the steepest UV slope among other results from various quasar samples \citep{Laor97,Vanden01,Telfer02,Stevans14,Selsing16}.
If we adopt this rather extreme slope of $-1.9$, the exponent in Eq. (\ref{equ:e912}) changes to 0.1. 
However, the UV emissivity increases only by a factor of 1.27. 
So, the maximal UV ionizing photon fraction changes from 45\% to 57\%---but this can be considered as only a modest increase.

\section{SUMMARY \label{sec:summary}}

We present the advanced results of our $z\sim5$ quasar survey with IMS \citepalias{Kim19}.
Our findings are as follows.

\begin{enumerate}
\item{
Based on the follow-up spectroscopy carried out with GMOS on the Gemini-South 8 m Telescope, we newly identified eight $z\sim5$ quasars in the IMS survey area. 
Using our high-redshift quasar model, we measured their redshifts and magnitudes in the ranges of $4.71\leq z \leq 5.15$ and $-26.2 \leq M_{1450} \leq -23.3$.
}
\item{
Considering the survey completeness with our selection criteria, we derived the binned and parametric $z\sim5$ QLFs.
The selection criteria are considered either with only broadband colors (case 1; 43 quasars) or broadband and medium-band colors (case 2; 32 quasars).
Including the SDSS bright quasar sample of \citetalias{Yang16}, the parametric QLFs in both cases are well determined without any fixed parameters.
}
\item{
We find a relatively flat faint-end slope of $-1.2$ compared to previous studies at $z\sim5$, although our 1$\sigma$ limit allows steeper slopes down to $\alpha\sim-2$.
We calculated a low ionizing emissivity of $\epsilon_{912}=5.0\times10^{23}$ erg s$^{-1}$ Hz$^{-1}$ Mpc$^{-3}$ and a number density of UV ionizing photons of $\dot{n}=4.0\times10^{49}$ Mpc$^{-3}$ s$^{-1}$.
This implies that the quasars do not produce all of the ionizing photons at $z\sim5$.
}
\end{enumerate}

Our $z\sim5$ QLF that is well defined at $M_{1450}<-23$ mag with a better parameter estimation than previous surveys, but our imaging survey depths limits us from determining the QLF at $M_{1450}>-23$ mag.
Hence, our conclusion regarding the contribution of quasars to the IGM-ionizing photons relies on the uncertain extrapolation of our QLF to $M_{1450}>-23$ mag.
The above conclusion may change significantly if there is an abrupt increase in the number of quasars at the faintest limit such as a large number of obscured quasars at high redshifts (e.g., \citealt{Aird15}).
Deeper imaging data are necessary to settle down this issue, which could come from the Legacy Survey of Space and Time (LSST) and future space telescope missions.
Reliable identification of these faint quasars may be challenging even with the future 30-m class telescopes.
But even in such a challenging regime, we expect that our medium-band technique will be able to identify faint high-redshift quasars if used on 4 to 8-m class telescopes.

\acknowledgments

This work was supported by the National Science Foundation of China (11721303, 11890693), the National Research Foundation of Korea (NRF) grant (2020R1A2C3011091) funded by the Korean government (MSIP),  K-GMT Science Program (PID:GS-2018B-Q-217, GS-2019A-Q-218) of Korea Astronomy and Space Science Institute (KASI), and the National Key R\&D Program of China (2016YFA0400703).
Y. K. acknowledges the support from the China Postdoc Science General (2020M670022) and Special (2020T130018) Grants funded by the China Postdoctoral Science Foundation.
S. L. acknowledges the support from the National Research Foundation of Korea (NRF) grant (2020R1I1A1A01060310) funded by the Korean government (MIST).
H. D. J. was supported by Basic Science Research Program through the National Research Foundation of Korea (NRF) funded by the Ministry of Education (NRF-2017R1A6A3A04005158).
We thank Jinyi Yang for providing the bright $z\sim5$ quasar sample and the completeness functions described in \cite{Yang16}.

This paper includes data obtained at the Gemini Observatory, acquired through the Gemini Science Archive, and processed using the Gemini IRAF package, which is operated by the Association of Universities for Research in Astronomy, Inc., under a cooperative agreement with the NSF on behalf of the Gemini partnership: the National Science Foundation (United States), the National Research Council (Canada), CONICYT (Chile), the Australian Research Council (Australia), Minist\'{e}rio da Ci\^{e}ncia, Tecnologia e Inova\c{c}\~{a}o (Brazil), and Ministerio de Ciencia, Tecnolog\'{i}a e Innovaci\'{o}n Productiva (Argentina).
This paper includes data obtained with MegaPrime/MegaCam, a joint project of CFHT and CEA/IRFU, at the Canada--France--Hawaii Telescope (CFHT), which is operated by the National Research Council (NRC) of Canada, the Institut National des Science de l'Univers of the Centre National de la Recherche Scientifique (CNRS) of France, and the University of Hawaii. 
This work is based in part on data products produced at Terapix available at the Canadian Astronomy Data Centre as part of the Canada--France--Hawaii Telescope Legacy Survey, a collaborative project of NRC and CNRS.
The United Kingdom Infrared Telescope (UKIRT) is supported by NASA and operated under an agreement between the University of Hawaii, the University of Arizona, and Lockheed Martin Advanced Technology Center; operations are enabled through the cooperation of the Joint Astronomy Centre of the Science and Technology Facilities Council of the U.K.
This paper includes data gathered with the 6.5 m Magellan telescopes located at Las Campanas Observatory, Chile.
This paper includes data taken at The McDonald Observatory of The University of Texas at Austin.
We would like to thank Editage (www.editage.co.kr) for English language editing.

\vspace{5mm}
\facilities{UKIRT (WFCAM), Struve (SQUEAN), Magellan:Baade (IMACS), Gemini:South (GMOS-S), CFHT (MegaCam), Sloan }

\begin{figure*}
\centering
\epsscale{1.0}
\plotone{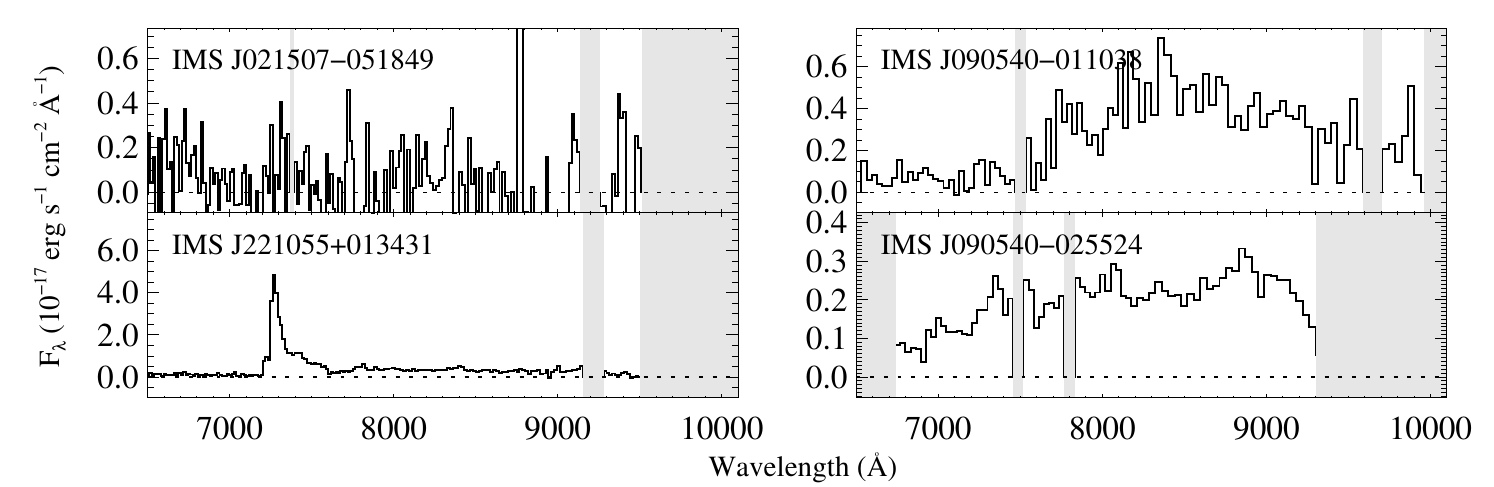}
\caption{
Left: IMACS optical spectra of two of the IMS $z\sim5$ quasars that were newly discovered in this work.
IMS J021507$-$051849 was not detected with IMACS.
Right: GMOS-S optical spectra of two non-quasars.
They have no Ly$\alpha$ break with continuum emissions.
The shaded regions represent the bad columns, hot pixels, CCD gaps, or wavelength ranges that are not covered by the observational configuration. 
\label{fig:appendix}}
\end{figure*}

\begin{deluxetable*}{lcccc}
\tabletypesize{\scriptsize}
\tablecaption{Spectroscopic Observations of Quasars and Non-quasars \label{tbl:appendix}}
\tablewidth{0pt}
\tablehead{
\colhead{ID} & \colhead{Telescope/Instrument} & \colhead{Date} & \colhead{Exposure time (s)} & \colhead{Seeing (arcsec)}
}
\startdata
\multicolumn{5}{c}{Confirmed $z\sim5$ Quasars}\\
IMS J021507$-$051849 & Magellan/IMACS 	& 2018 Sep 10	& 1200	& 0.8 \\
IMS J221055$+$013430 & Magellan/IMACS 	& 2018 Sep 9	& 1800	& 0.8 \\
\hline
\multicolumn{5}{c}{Nonquasars}\\
IMS J090540$-$011038 	& Gemini/GMOS		& 2019 Jan 13	& 605	& 1.1 \\
IMS J090540$-$025524	& Gemini/GMOS 	& 2019 Feb 11	& 4356	& 1.1 \\
\enddata
\end{deluxetable*}

\clearpage

\appendix

\begin{figure*}
\centering
\epsscale{0.7}
\plotone{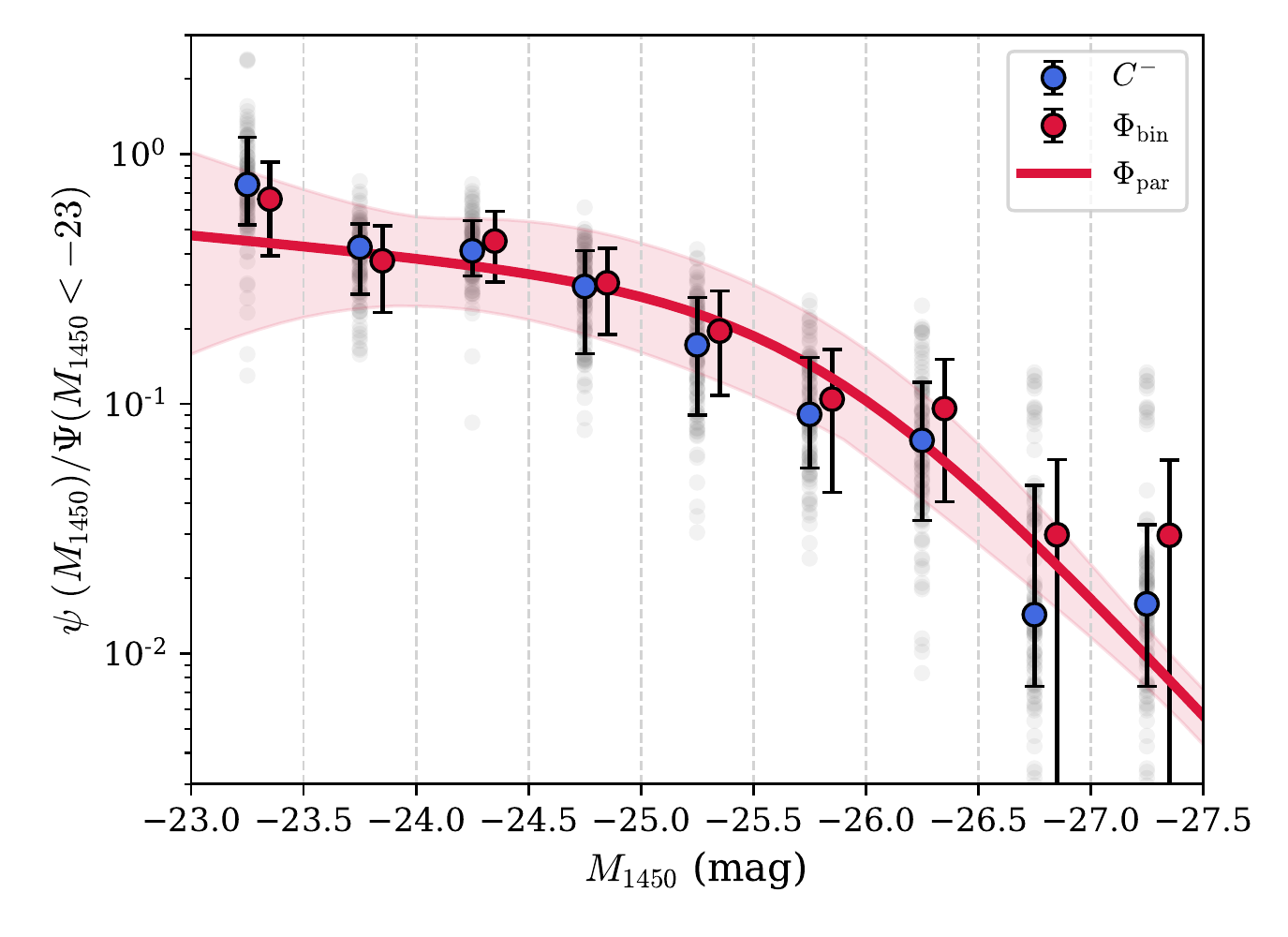}
\caption{
Nonparametric QLF derived using the Lynden-Bell's $C^{-}$ estimator \citep{Lynden71}.
The gray circles denote the marginal distributions of $\psi(M_{1450})$ in each magnitude bin from the 100 bootstrap sample, and their median values with 1$\sigma$ errors (16\% and 84\% percentile values) are shown as the blue circles with error bars.
The red circles denote the binned QLF for the case 1 sample, shifted horizontally by $-0.1$ mag for easy comparison.
The parametric QLF with its 1$\sigma$ confidence level (case 1) is shown as the red solid line with the shaded area.
All the QLFs are normalized by the total number of quasars expected at $M_{1450}<-23$ mag.
The vertical dashed lines indicate the magnitude bins' edges.
\label{fig:cminus}}
\end{figure*}

\section{ADDITIONAL SPECTROSCOPY RESULTS\label{sec:addspec}}

Here, we describe the optical spectra of our quasars/candidates, which are not included in the analysis in this paper.
The observations are summarized in Table \ref{tbl:appendix}, and their spectra are shown in Figure \ref{fig:appendix}.

Of the eight newly discovered $z\sim5$ quasars, IMS J021507$-$051849 and IMS J221054$+$013430 were also observed with the Inamori--Magellan Areal Camera and Spectrograph (IMACS; \citealt{Dressler11}) on the Magellan Baade 6.5 m Telescope.
The observing configuration was set to follow that used in \citetalias{Kim19}: using a grating of 150 lines mm$^{-1}$ with a resolution of $R\sim600$, OG570 filters to avoid a zeroth-order overlap, and spectral/spatial binning of $2\times2$.
We followed the general steps for spectral reduction, except for the flux calibration with standard stars.
Instead, we scaled the fluxes with their $i'$-band magnitudes.
These are excluded in the main text because the GMOS with which our fiducial spectra were obtained is more sensitive than IMACS.

We also present the two candidates that were identified as non-quasars with GMOS-S: IMS J090540$-$011038 and IMS J090540$-$025524.
The observing configurations were as described in Section \ref{sec:newquasars} and in \citetalias{Kim19}.
In their 1D and 2D spectra, there is no Ly$\alpha$ break but continuum emission across the whole wavelength range, meaning that they are not high-redshift quasars.

\section{NONPARAMETRIC DETERMINATION OF QLF}\label{sec:cminus}

The QLFs determined in the main text are based on the small number statistics, so the methods we used could affect the exact QLF estimations.
For instance, the $1/V_{\rm max}$ method critically depends on how we bin the sample especially in the case with a handful of quasars.
In this section, we describe the nonparametric determination of QLF, using the Lynden-Bell's $C^{-}$ method \citep{Lynden71} that requires no assumption on the distribution for the one-side-truncated data sets.
Thus, it is suitable for our quasar samples and the survey limits; the brightest quasar in our sample (IMS J220107$+$030208; $i'=19.1$ mag) is much fainter than the saturation limit of CFHTLS ($i'\sim17$ mag).

QLF is canonically described in the form of a bivariate function of magnitude and redshift, e.g., $\Phi(M,z)$.
However, our IMS quasars are in the narrow redshift range of $4.7<z<5.4$, so what we actually derived in Section \ref{sec:binpar} is similar to the univariate function of magnitude at the specific redshift of $z=5$.
Hence, in this narrow redshift range, the marginal distribution of the bivariate QLF in the magnitude direction, $\psi(M)$, can be approximated by $\psi(M)\approx \Phi(M,z)$.
Note that we performed the standard correlation test between the magnitude and redshift of our sample with the Kendall's $\tau$ statistic (see details in \citealt{Efron92,Fan01b,Schindler18,Schindler19}), giving $\tau=0.04$ which means they are uncorrelated at $\sim1\sigma$ level.

With the Lynden-Bell's $C^{-}$ estimator \citep{Lynden71}, the cumulative luminosity function $\Psi(M)=\int \psi(M) dM$ can be recovered:

\begin{equation}
\Psi(M_{j}) = \Psi(M_1) \prod_{i=2}^{j} \left( 1 + \frac{1}{C^{-}(M_i)} \right) \label{equ:psi},
\end{equation}

\noindent for the sample sorted by magnitude ($M_1<...<M_{i-1}<M_i<...<M_{N}$).
The estimator $C^{-}(M_i)$ indicates the total number of object of which magnitudes are brighter than $M_i$ (but the object $i$ itself is omitted).
But the problem is that this direct usage of $C^{-}(M_i)$ can be applied to a sample with a sharp boundary for selection, while our selection probability has a smooth boundary in the magnitude direction.
Instead of $C^{-}(M_i)$, \cite{Fan01} generalized the total \textit{weighted} number of quasars $N_i$ for arbitrary selection functions  (see also \citealt{Schindler18,Schindler19}).
Given a selection probability function $p(M,z)$, one can construct the quantity $N_i$:

\begin{equation}
N_{i} = \sum_{j} \frac{p(M_i,z_j)}{p(M_j,z_j)},
\end{equation}

\noindent where the sum extends over $M_j<M_i$.
Using the quantity $N_i$ instead of $C^{-}(M_i)$ in equation (\ref{equ:psi}), we evaluate $\Psi(M_{1450})$ for the 43 IMS quasars (case 1).
Considering the discontinuity of $\Psi(M_{1450})$, we calculate $\psi(M_{1450})=\Delta \Psi / \Delta M_{1450}$ in the magnitude bins whose size $\Delta M_{1450}$ is identical to that of the binned QLF in Section \ref{sec:binpar}.
Given the edges of each bin, ($M_l$, $M_u$), where $M_u-M_l=\Delta M_{1450}$, we evaluate $\Delta \Psi = \Psi(M_u)-\Psi(M_l)$ using the linear interpolation on $\Psi(M_{1450})$ at the edge magnitudes.
We note that the $\psi(M_{1450})$ of the faintest bin ($-23.5\leq M_{1450} <-23.0$) could be overestimated in this linear interpolation process, because the faintest object in our sample has $M_{1450}=-23.3$ mag, brighter than the bin's edge.

Figure \ref{fig:cminus} shows the $\psi(M_{1450})$ of the IMS quasars from the 100 bootstrap samples.
While the $C^{-}$ method can recover the shape of the marginal luminosity function, it does not provide any information related to its normalization.
Therefore, the marginal distribution is normalized by the total number of quasars expected,  $\Psi(M_{1450}<-23)=\int^{-23}_{-\infty} \psi(M_{1450}) dM_{1450}$.
We also plot our binned and parametric QLFs, also normalized by the cumulative luminosity function of the parametric QLF at $M_{1450}<-23$ mag.
They are consistent with the marginal distribution $\psi(M_{1450})$ within 1$\sigma$ errors.
Therefore, we can confidently say that our binning for the binned QLF is reliable, and the fitting for the parametric QLF as well.


\begin{thebibliography}{}


\bibitem[Aird et al.(2015)]{Aird15} Aird, J., Coil, A.~L., Georgakakis, A., et al.\ 2015, \mnras, 451, 1892

\bibitem[Akiyama et al.(2018)]{Akiyama18} Akiyama, M., He, W., Ikeda, H., et al.\ 2018, \pasj, 70, S34

\bibitem[Avni, \& Bahcall(1980)]{Avni80} Avni, Y., \& Bahcall, J.~N.\ 1980, \apj, 235, 694

\bibitem[Ba{\~n}ados et al.(2014)]{Banados14} Ba{\~n}ados, E., 
Venemans, B.~P., Morganson, E., et al.\ 2014, \aj, 148, 14 

\bibitem[Ba{\~n}ados et al.(2016)]{Banados16} Ba{\~n}ados, E., 
Venemans, B.~P., Decarli, R., et al.\ 2016, \apjs, 227, 11 

\bibitem[Ba{\~n}ados et al.(2018)]{Banados18} Ba{\~n}ados, E., Venemans, B.~P., Mazzucchelli, C., et al.\ 2018, \nat, 553, 473 

\bibitem[Becker \& Bolton(2013)]{Becker13} Becker, G.~D., \& Bolton, J.~S.\ 2013, \mnras, 436, 1023

\bibitem[Bertin 
\& Arnouts(1996)]{Bertin96} Bertin, E., \& Arnouts, S.\ 1996, \aaps, 117, 393 

\bibitem[Bolton, \& Haehnelt(2007)]{Bolton07} Bolton, J.~S., \& Haehnelt, M.~G.\ 2007, \mnras, 382, 325

\bibitem[Boutsia et al.(2018)]{Boutsia18} Boutsia, K., Grazian, A., Giallongo, E., et al.\ 2018, \apj, 869, 20

\bibitem[Casali et al.(2007)]{Casali07} Casali, M., Adamson, A., Alves de Oliveira, C., et al.\ 2007, \aap, 467, 777

\bibitem[Choi et al.(2015)]{Choi15} Choi, N., Park, W.-K., Lee, H.-I., et al.\ 2015, Journal of Korean Astronomical Society, 48, 177

\bibitem[D'Aloisio et al.(2018)]{Daloisio18} D'Aloisio, A., McQuinn, M., Davies, F.~B., et al.\ 2018, \mnras, 473, 560


\bibitem[D'Aloisio et al.(2020)]{Daloisio20} D'Aloisio, A., McQuinn, M., Trac, H., et al.\ 2020, arXiv e-prints, arXiv:2002.02467


\bibitem[Davies et al.(2018)]{Davies18} Davies, F.~B., Hennawi, J.~F., Ba{\~n}ados, E., et al.\ 2018, \apj, 864, 142

\bibitem[Dietrich et al.(2002)]{Dietrich02} Dietrich, M., Hamann, F., Shields, J.~C., et al.\ 2002, \apj, 581, 912 

\bibitem[Diamond-Stanic et al.(2009)]{Diamond09} Diamond-Stanic, A.~M., Fan, X., Brandt, W.~N., et al.\ 2009, \apj, 699, 782

\bibitem[Dressler et al.(2011)]{Dressler11} Dressler, A., Bigelow, B., Hare, T., et al.\ 2011, \pasp, 123, 288 

\bibitem[Efron \& Petrosian(1992)]{Efron92} Efron, B., \& Petrosian, V.\ 1992, \apj, 399, 345

\bibitem[Eilers et al.(2017)]{Eilers17} Eilers, A.-C., Davies, F.~B., Hennawi, J.~F., et al.\ 2017, \apj, 840, 24

\bibitem[Fan et al.(2006)]{Fan06} Fan, X., Strauss, M.~A., 
Becker, R.~H., et al.\ 2006, \aj, 132, 117 

\bibitem[Fan et al.(2001a)]{Fan01} Fan, X.,
 Narayanan, V.~K., Lupton, R.~H., et al.\ 2001a, \aj, 122, 2833 

\bibitem[Fan et al.(2001b)]{Fan01b} Fan, X., Strauss, M.~A., Schneider, D.~P., et al.\ 2001b, \aj, 121, 54

\bibitem[Fleming et al.(1995)]{Fleming95} Fleming, D.~E.~B., Harris, W.~E., Pritchet, C.~J., et al.\ 1995, \aj, 109, 1044

\bibitem[Giallongo et 
al.(2015)]{Giallongo15} Giallongo, E., Grazian, A., Fiore, F., et al.\ 2015, \aap, 578, A83 

\bibitem[Giallongo et al.(2019)]{Giallongo19} Giallongo, E., Grazian, A., Fiore, F., et al.\ 2019, \apj, 884, 19

\bibitem[Glikman et al.(2011)]{Glikman11} Glikman, E., Djorgovski, S.~G., Stern, D., et al.\ 2011, \apjl, 728, L26

\bibitem[Grazian et al.(2018)]{Grazian18} Grazian, A., Giallongo, E., Boutsia, K., et al.\ 2018, \aap, 613, A44

\bibitem[Grazian et al.(2020)]{Grazian20} Grazian, A., Giallongo, E., Fiore, F., et al.\ 2020, arXiv e-prints, arXiv:2006.02451

\bibitem[Haardt \& Madau(2012)]{Haardt12} Haardt, F., \& Madau, P.\ 2012, \apj, 746, 125 

\bibitem[Hook et al.(2004)]{Hook04} Hook, I.~M., J{\o}rgensen, I., Allington-Smith, J.~R., et al.\ 2004, \pasp, 116, 425 

\bibitem[Hudelot et al.(2012)]{Hudelot12} Hudelot, P., 
Cuillandre, J.-C., Withington, K., et al.\ 2012, VizieR Online Data 
Catalog, 2317, 0 

\bibitem[Im et al.(1997)]{Im97} Im, M., Griffiths, R.~E., 
\& Ratnatunga, K.~U.\ 1997, \apj, 475, 457 

\bibitem[Jiang et al.(2016)]{Jiang16} Jiang, L., 
McGreer, I.~D., Fan, X., et al.\ 2016, \apj, 833, 222 

\bibitem[Jeeson-Daniel et al.(2014)]{Jeeson14} Jeeson-Daniel, A., Ciardi, B., \& Graziani, L.\ 2014, \mnras, 443, 2722

\bibitem[Jeon et al.(2017)]{Jeon17} Jeon, Y., Im, M., Kim, D., et al.\ 2017, \apjs, 231, 16 

\bibitem[Jeon et al.(2016)]{Jeon16} Jeon, Y., Im, M., Pak, S., et al.\ 2016, Journal of Korean Astronomical Society, 49, 25 

\bibitem[Kashikawa et al.(2015)]{Kashikawa15} Kashikawa, N., 
Ishizaki, Y., Willott, C.~J., et al.\ 2015, \apj, 798, 28 

\bibitem[Kim et al.(2015)]{Kim15} Kim, Y., Im, M.,
Jeon, Y., et al.\ 2015, \apjl, 813, L35 

\bibitem[Kim et al.(2018)]{Kim18} Kim, Y., Im, M.,
Jeon, Y., et al.\ 2018, \apj, 855, 138

\bibitem[Kim et al.(2019)]{Kim19} Kim, Y., Im, M., Jeon, Y., et al.\ 2019, \apj, 870, 86

\bibitem[Kim et al.(2016)]{Kim16} Kim, S., Jeon, Y., Lee, H.-I., et al.\ 2016, \pasp, 128, 115004 

\bibitem[Kulkarni et al.(2019)]{Kulkarni19} Kulkarni, G., Worseck, G., \& Hennawi, J.~F.\ 2019, \mnras, 488, 1035

\bibitem[Lampton et al.(1976)]{Lampton76} Lampton, M., Margon, B., \& Bowyer, S.\ 1976, \apj, 208, 177

\bibitem[Laor et al.(1997)]{Laor97} Laor, A., Fiore, F., Elvis, M., et al.\ 1997, \apj, 477, 93

\bibitem[Lawrence et al.(2007)]{Lawrence07} Lawrence, A., Warren, S.~J., Almaini, O., et al.\ 2007, \mnras, 379, 1599 

\bibitem[Lusso et al.(2015)]{Lusso15} Lusso, E., Worseck, G., Hennawi, J.~F., et al.\ 2015, \mnras, 449, 4204

\bibitem[Lynden-Bell(1971)]{Lynden71} Lynden-Bell, D.\ 1971, \mnras, 155, 95

\bibitem[Madau et al.(1996)]{Madau96} Madau, P., Ferguson, H.~C., Dickinson, M.~E., et al.\ 1996, \mnras, 283, 1388 

\bibitem[Madau et al.(1999)]{Madau99} Madau, P., Haardt, F., \& Rees, M.~J.\ 1999, \apj, 514, 648

\bibitem[Madau(2017)]{Madau17} Madau, P.\ 2017, \apj, 851, 50

\bibitem[Marshall et al.(1983)]{Marshall83} Marshall, H.~L., Tananbaum, H., Avni, Y., et al.\ 1983, \apj, 269, 35

\bibitem[Mason et al.(2018)]{Mason18} Mason, C.~A., Treu, T., Dijkstra, M., et al.\ 2018, \apj, 856, 2

\bibitem[Mason et al.(2019)]{Mason19} Mason, C.~A., Naidu, R.~P., Tacchella, S., et al.\ 2019, \mnras, 489, 2669

\bibitem[Matsuoka et al.(2016)]{Matsuoka16} Matsuoka, Y., 
Onoue, M., Kashikawa, N., et al.\ 2016, \apj, 828, 26 

\bibitem[Matsuoka et al.(2018a)]{Matsuoka18a} Matsuoka, Y., Onoue, M., Kashikawa, N., et al.\ 2018a, \pasj, 70, S35

\bibitem[Matsuoka et al.(2018b)]{Matsuoka18b} Matsuoka, Y., Strauss, M.~A., Kashikawa, N., et al.\ 2018b, \apj, 869, 150

\bibitem[Matsuoka et al.(2019)]{Matsuoka19} Matsuoka, Y., Onoue, M., Kashikawa, N., et al.\ 2019, \apjl, 872, L2

\bibitem[Mazzucchelli et al.(2017)]{Mazzucchelli17} Mazzucchelli, C., Ba{\~n}ados, E., Venemans, B.~P., et al.\ 2017, \apj, 849, 91 

\bibitem[McGreer et al.(2013)]{McGreer13} McGreer, I.~D., Jiang, L., Fan, X., et al.\ 2013, \apj, 768, 105 

\bibitem[McGreer et al.(2018)]{McGreer18} McGreer, I.~D., Fan, X., Jiang, L., \& Cai, Z.\ 2018, \aj, 155, 131 

\bibitem[McGreer et al.(2015)]{McGreer15} McGreer, I.~D., Mesinger, A., \& D'Odorico, V.\ 2015, \mnras, 447, 499

\bibitem[Meiksin(2005)]{Meiksin05} Meiksin, A.\ 2005, \mnras, 356, 596

\bibitem[Mortlock et al.(2011)]{Mortlock11} Mortlock, D.~J., 
Warren, S.~J., Venemans, B.~P., et al.\ 2011, \nat, 474, 616 

\bibitem[Onoue et al.(2017)]{Onoue17} Onoue, M., Kashikawa, N., Willott, C.~J., et al.\ 2017, \apjl, 847, L15 

\bibitem[Parsa et al.(2018)]{Parsa18} Parsa, S., Dunlop, J.~S., \& McLure, R.~J.\ 2018, \mnras, 474, 2904 

\bibitem[Planck Collaboration et al.(2016)]{Planck16} Planck Collaboration, Adam, R., Aghanim, N., et al.\ 2016, \aap, 596, A108 

\bibitem[Planck Collaboration et al.(2018)]{Planck18} Planck Collaboration, Aghanim, N., Akrami, Y., et al.\ 2018, arXiv e-prints, arXiv:1807.06209

\bibitem[Reed et al.(2017)]{Reed17} Reed, S.~L., McMahon, R.~G., Martini, P., et al.\ 2017, \mnras, 468, 4702 

\bibitem[Ricci et al.(2017)]{Ricci17} Ricci, F., Marchesi, S., Shankar, F., La Franca, F., \& Civano, F.\ 2017, \mnras, 465, 1915 

\bibitem[Schindler et al.(2018)]{Schindler18} Schindler, J.-T., Fan, X., McGreer, I.~D., et al.\ 2018, \apj, 863, 144

\bibitem[Schindler et al.(2019)]{Schindler19} Schindler, J.-T., Fan, X., McGreer, I.~D., et al.\ 2019, \apj, 871, 258

\bibitem[Selsing et al.(2016)]{Selsing16} Selsing, J., Fynbo, J.~P.~U., Christensen, L., et al.\ 2016, \aap, 585, A87

\bibitem[Shen et al.(2011)]{Shen11} Shen, Y., Richards, G.~T., Strauss, M.~A., et al.\ 2011, \apjs, 194, 45

\bibitem[Shen et al.(2020)]{Shen20} Shen, X., Hopkins, P.~F., Faucher-Gigu{\`e}re, C.-A., et al.\ 2020, \mnras, doi:10.1093/mnras/staa1381


\bibitem[Shin et al.(2020)]{Shin20} Shin, S., Im, M., Kim, Y., et al.\ 2020, \apj, 893, 45

\bibitem[Shull et al.(2012)]{Shull12} Shull, J.~M., Harness, A., Trenti, M., et al.\ 2012, \apj, 747, 100

\bibitem[Stevans et al.(2014)]{Stevans14} Stevans, M.~L., Shull, J.~M., Danforth, C.~W., et al.\ 2014, \apj, 794, 75

\bibitem[Telfer et al.(2002)]{Telfer02} Telfer, R.~C., Zheng, W., Kriss, G.~A., et al.\ 2002, \apj, 565, 773

\bibitem[Trakhtenbrot et al.(2011)]{Trakhtenbrot11} Trakhtenbrot, B., Netzer, H., Lira, P., \& Shemmer, O.\ 2011, \apj, 730, 7 

\bibitem[Vanden Berk et al.(2001)]{Vanden01} Vanden Berk, D.~E., 
Richards, G.~T., Bauer, A., et al.\ 2001, \aj, 122, 549 

\bibitem[Venemans et al.(2013)]{Venemans13} Venemans, B.~P., 
Findlay, J.~R., Sutherland, W.~J., et al.\ 2013, \apj, 779, 24 

\bibitem[Venemans et al.(2015a)]{Venemans15a} Venemans, B.~P., 
Ba{\~n}ados, E., Decarli, R., et al.\ 2015a, \apjl, 801, L11 

\bibitem[Venemans et al.(2015b)]{Venemans15b} Venemans, B.~P., 
Verdoes Kleijn, G.~A., Mwebaze, J., et al.\ 2015b, \mnras, 453, 2259 

\bibitem[Vito et al.(2018)]{Vito18} Vito, F., Brandt, W.~N., Yang, G., et al.\ 2018, \mnras, 473, 2378


\bibitem[Wang et al.(2016)]{Wang16} Wang, F., Wu, X.-B., Fan, X., et al.\ 2016, \apj, 819, 24 

\bibitem[Wang et al.(2017)]{Wang17} Wang, F., Fan, X., Yang, J., et al.\ 2017, \apj, 839, 27

\bibitem[Wang et al.(2018)]{Wang18} Wang, F., Yang, J., Fan, X., et al.\ 2018, \apjl, 869, L9

\bibitem[Wang et al.(2019)]{Wang19} Wang, F., Yang, J., Fan, X., et al.\ 2019, \apj, 884, 30

\bibitem[Willott et al.(2010)]{Willott10b} Willott, C.~J., 
Delorme, P., Reyl{\'e}, C., et al.\ 2010, \aj, 139, 906 

\bibitem[Wu et al.(2015)]{Wu15} Wu, X.-B., Wang, F., Fan, 
X., et al.\ 2015, \nat, 518, 512 

\bibitem[Wyithe \& Bolton(2011)]{Wyithe11} Wyithe, J.~S.~B., \& Bolton, J.~S.\ 2011, \mnras, 412, 1926

\bibitem[Yang et al.(2016)]{Yang16} Yang, J., Wang, F., Wu, X.-B., et al.\ 2016, \apj, 829, 33 

\bibitem[Yang et al.(2017)]{Yang17} Yang, J., Fan, X., Wu, X.-B., et al.\ 2017, \aj, 153, 184 

\bibitem[Yang et al.(2019a)]{Yang19a} Yang, J., Wang, F., Fan, X., et al.\ 2019a, \apj, 871, 199

\bibitem[Yang et al.(2019b)]{Yang19b} Yang, J., Wang, F., Fan, X., et al.\ 2019b, \aj, 157, 236

\bibitem[Yang et al.(2020)]{Yang20} Yang, J., Wang, F., Fan, X., et al.\ 2020, \apjl, 897, L14


\end{thebibliography}
\end{document}